\documentclass[journal]{IEEEtran}
\pdfoutput=1
\ifCLASSINFOpdf

\else

\fi
\usepackage{amsmath,amssymb,amsfonts}
\usepackage{cite}
\usepackage{algorithm}  
\usepackage{algorithmicx}  
\usepackage{algpseudocode}
\usepackage{textcomp}
\usepackage{longtable,booktabs}
\usepackage{lscape}
\usepackage{graphicx}
\usepackage{caption}
\usepackage{subcaption}
\usepackage{tabularx}
\usepackage{array}
\usepackage{algpseudocode}
\usepackage{makecell}
\usepackage{xcolor}

\begin{document}

\title{A Lightweight and Scalable Design of Segment Routing  in Broadband LEO Constellations Using Landmark-Based Skeleton Graphs}
\author{Menglan Hu, Chenxin Wang, Bin Cao, Benkuan Zhou, Yan Dong, Kai Peng}
\maketitle
\begin{abstract}
Emerging Low Earth Orbit (LEO) broadband constellations hold significant potential to provide advanced Internet services due to inherent geometric features of the grid topology. However, high dynamics, unstable topology changes, and frequent route updates bring significant challenge to fast and adaptive routing policies. In addition, since computing, bandwidth, and storage resources in each LEO satellite is strictly limited, traffic demands are typically unbalanced, further enlarging the challenge to scalable routing policies with load balancing. Nevertheless, most existing research failed to address the above difficulties. Therefore, this paper proposes a lightweight and scalable protocol of segment routing through landmark-based skeleton graphs. To improve the overall performance, we design an efficient multipath segment routing algorithm. First, the algorithm partitions the network into multiple regions to construct skeleton paths, which can effectively guide packet forwarding and reduce the operating costs. In each region, multipath probabilistic routing is used to achieve uniform traffic distribution, avoiding hotspot congestion. Furthermore, the flexible hierarchical partitioning and localized segmented routing is employed for fine-grained traffic control and QoS guarantee combined with adaptive local single-path routing. Finally, experimental results validate our method's superior performance in terms of response time and network utility.
\end{abstract}
\IEEEpeerreviewmaketitle
\section{Introduction}
The rapid expansion of satellite communication technologies, especially Low Earth Orbit (LEO) satellite constellations, has introduced unique opportunities for global internet coverage. Satellite communications operators, spearheaded by Starlink \cite{starlink} and Oneweb \cite{oneweb}, are implementing the deployment of satellites in low-orbit constellations with the objective of establishing global communications networks. These constellations promise high-speed, low-latency internet access across diverse and extensive geographical areas, including remote and polar regions, thereby addressing global connectivity disparities. Compared to traditional terrestrial networks, LEO systems benefit from the regular grid topology which allows for direct use of coordinate and positional information for routing. The inherent geometric features of the grid topology can be easily exploited to design efficient routing and traffic management methods, making it suitable for geometric and geographic routing. A homogeneous network managed by a single operator is also suitable for centralized control and optimization, making it ideal for segmented and source routing techniques.

Simultaneously, the advancement of Software-Defined Networking (SDN) application in satellite communications is attracting considerable interest. The integration of SDN into the satellite domain and the construction of an integrated space and terrestrial networking protocol suite are driven by the objective of extending advanced terrestrial Internet technologies, with the aim of achieving flexible programmability of path selections \cite{zheng_sdn_2024}.  This encourages us to investigated the implementation of traffic engineering (TE) schemes in software-defined enhanced satellite networks. A few previous studies in SDN investigate the centralized routing and distributed routing strategies to satisfy the minimum end-to-end hop-count and queuing delay constraints \cite{li_stigmergy_2024}. Such work made no efforts to optimize the overall bandwidth consumption and the overhead cost of routing information. In this case,  this paper employs a segment routing strategy based on skeleton graphs, which is more suitable to exploit SDN structure to minimize the communication delay and routing computation time.

Besides, current satellite routing algorithms struggle to handle the high-density, high-mobility nature of LEO constellations efficiently due to the rapid topology changes and varying link states, leading to frequent route recalculations and updates that impose a heavy computational and communication overhead. Consequently, for low-orbit constellations with constrained resources, a lightweight and scalable load-balancing strategy is imperative due to the limited bandwidth and inadequate computation capabilities of a single satellite. Existing research overlooks the advantages of LEO satellite networks and fails to address these significant challenges effectively. Reference \cite{mao_intelligent_2024} predominantly focusing on Quality of Service (QoS) constraints, with less emphasis on the network-wide traffic management strategy. Hence, there is a highly valuable to employ the load-balancing protocol in the emerging massive LEO constellations to enhance the satellite communication effectiveness.

However, in comparison to conventional terrestrial networks, LEO satellite internet also encounters considerable obstacles. Satellite Internet usually contains tens of thousands of satellites, each of which can play the role of a router. Routing and traffic management for a large number of satellites is difficult and requires high algorithm scalability. In addition, the number of satellites in the LEO constellation is considerable, but the laser link bandwidth of these satellites is relatively limited. This can result in the occurrence of network performance bottlenecks to occur in the face of large amounts of network traffic. Previous work on load-balancing satellite network proposed a rectangular transmission region between sources and destinations to control the delay and transmission cost \cite{deng_distance-based_2023}, which are unable to regulate network-wide traffic trends. 

To address the above needs and challenges, this paper designs a lightweight and scalable segmented routing algorithm within the SDN paradigm that fully leverages the inherent geometric characteristics and positional information of the grid topology, aiming for a flexible, adaptive, and highly load-balanced solution. Our contributions are multi-fold:

\begin{enumerate}
    \item We propose a lightweight source routing strategy based on skeleton paths. By partitioning the network into sub-regions and abstracting it into a skeletal graph, long-distance multi-hop inter-satellite routes are converted into short skeletal paths. A minimal amount of skeletal path information can effectively direct the actual packet forwarding through multiple landmarks, thus avoiding hot spots and congested areas. In each underlying segment, lightweight fuzzy routing is achieved through multipath probabilistic forwarding, distributing tunnel traffic across a strip-like region. This design dramatically reduces the cost of interstellar routing computation, updating, and storage, enabling fast and  adaptable traffic engineering.
    \item Additionally, this paper presents a hierarchical partitioning and localized segmented routing algorithm to support specific QoS requirements for different services. By further dividing the partition into subareas, a segment can be split into subsegments. The combination of probabilistic routing with local table routing provides a robust solution that is highly adaptable and resilient to frequent link failures. This allows for the generation of a list of subsegments and the implementation of more detailed local segment routing.  
    \item Moreover, we demonstrate through formula derivation and experimental evaluation that the proposed routing policy is close to the benchmark algorithm in terms of routing performance with a much lower computational overhead, thereby reducing considerable execution time.
\end{enumerate}

The reminder of this paper is organized as follows. Section II introduces related work. Section III presents mathematical models. Section IV describes multi-path segmented routing protocols. Section V proposes hierarchical segment routing schemes. Section VI presents simulation results to evaluate the different routing strategies, with conclusions following in Section VII.

\section{Related work}
As the 6G era approaches, low Earth orbit satellite networks are seen as the crucial infrastructure for achieving true global coverage.
Ma demonstrated that Starlink is an effective solution for achieving ubiquitous internet coverage on Earth. They also analyzed the relay strategies and limitations of Starlink's connectivity.\cite{ma2023network}

Liang conducted a study on the effects of varying laser inter-satellite link ranges on satellite transmission power and network latency.\cite{liang2023free} Their findings indicate that an increase in inter-satellite link distance results in reduced average network latency, accompanied by a corresponding rise in the average satellite transmission power.

To achieve seamless global satellite coverage, an increasing number of satellites are being deployed, making the allocation of satellite network resources increasingly complex. As an efficient and highly adaptable network architecture, SDN offers a novel solution for the flexible configuration and management of LEO satellite networks and has been widely applied to address satellite network-related challenges. Qi et al. \cite{qi2022sdn} proposed an SDN-based multipath routing strategy that enhances transmission efficiency by dynamically sensing network conditions and the load on switching nodes within the network. Guo et al. \cite{guo2022static} proposed a Static and Dynamic Allocation (SPDA) method to address the Controller Placement Problem (CPP) in LEO satellite networks. Experimental results demonstrate that SPDA not only reduces switch-controller latency more effectively than existing methods but also achieves better load balancing performance. Papa et al. \cite{papa2020design} proposed an SDN architecture that quantifies reconfiguration and migration cost parameters to minimize the overhead imposed on the network.

The dynamic changes in network topology pose significant challenges to the management of LEO systems. The development and prediction of topology models have become key technologies for addressing these challenges.
Chen et al. \cite{chen2023gcn} proposed a prediction method based on Graph Convolutional Networks and Gated Recurrent Units, which not only predicts link changes in LEO constellations simultaneously but also significantly reduces memory consumption and computation time. Lai et al. \cite{lai2023achieving} introduced an innovative network model that addresses topology uncertainty by transforming topology changes caused by different failures into traffic fluctuations. Han et al. \cite{han2022time} proposed a weighted spatiotemporal evolution graph based on link attributes to model the time-varying topology of LEO satellite networks, with the goal of improving the adaptability of satellite routing.

The highly dynamic and time-varying topology, along with the limited onboard processing capabilities, imposes stringent demands on routing schemes. In recent years, an increasing number of researchers have focused on optimizing routing algorithms to address these challenges. F. Akyildiz et al. \cite{akyildiz2003distributed} proposed a distributed multicast routing scheme for multi-layer satellite IP networks, aiming to minimize the cost of multicast trees within the satellite network. Wang et al. \cite{wang2023intelligent} developed an intelligent region-based hybrid routing method that combines centralized and distributed approaches, designed to dynamically adapt to congestion-prone areas. Liu et al. \cite{liu2023drl} proposed a deep reinforcement learning (DRL)-powered intelligent routing algorithm capable of efficiently managing multiple subflow groups.
Pi et al. \cite{pi2022dynamic} proposes a dynamic Inter-plane Inter-satellite Links Planning method based on Multi-Agent deep reinforcement learning (MA-IILP) to optimize the total throughput and interplane ISL switching rate.

In recent studies, regional network segmentation has been proposed to reduce network complexity and enhance routing efficiency. Li et al. \cite{li2023leo} proposed a method that divides the LEO mega-constellation into regions with four quadrants, enabling multipath routing and forwarding through periodic link state updates instead of global computation. This approach improves throughput while effectively reducing packet loss. Mao et al. \cite{mao2024intelligent} employed regional network segmentation to reduce the complexity of routing.
\begin{figure}[t]
\centering
      \includegraphics[scale=0.42]{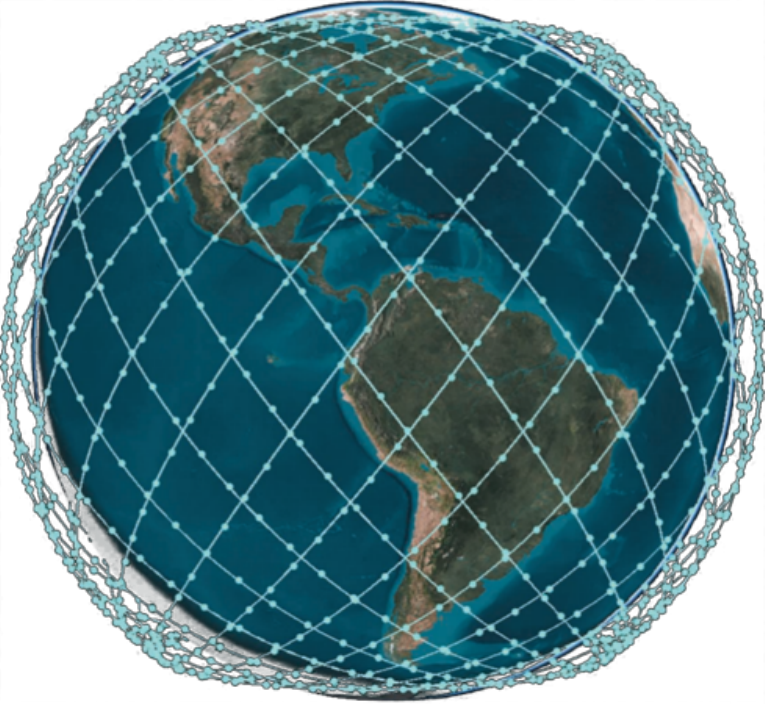}
\caption{Illustration of the Walker Delta Constellation}
\label{partition}
\end{figure}
\section{Model}

As illustrated in Fig. 1, this paper considers the Walker Delta constellation, which is designed to cover areas with a higher density of users. This constellation comprises $W_p$ orbital planes of inclination $W_i$, each with $W_N$ satellites distributed uniformly. The constellation orbital phase factor is defined as $W_f\in[0, W_p-1]$, thus the angle between the neighbouring satellites can be calculated as $\Delta\omega=2\pi W_f$/$W_p$ * $W_N$.

Establishing stable inter-satellite laser communication links is challenging due to the rapid relative motion between satellites during uplink and downlink operations. To address this, our approach concentrates on inter-orbit ISLs formed within isotropic orbital planes and stable intra-orbit ISLs. Each satellite connects to four neighboring satellites through ISLs, creating a $t \times p$ 2D network with a torus topology, represented as  $G(V, E)$. Satellites are identified by coordinates $(x, y)$, while the connections between them are depicted as edges $e$. Under normal conditions, these edges maintain consistent bandwidth capabilities.

Moreover, emerging Internet Service Providers (ISPs) have begun utilizing inter-satellite laser links in lieu of conventional microwave links for low-latency inter-satellite communication. Consequently, we elucidate the impact of distance between neighbouring satellites on latency, while solely focusing on the bandwidth consumption and utilization of the links to measure the performance of the algorithms in terms of load balancing and traffic engineering.

To address the impact of the dynamics of LEO satellites on routing computation, this paper employs time-slicing techniques to achieve agile inter-satellite local routing and network-wide traffic regulation. By discretely dividing the time domain into time slices, the network topology can be regarded as static in each time slice. Given the knowledge of the rules governing the satellite operation, including its orbit and speed, the network topology is fixed and can be calculated in advance for each short time slot.

It is worth noting that although the LEO satellites move at high speeds and the service area is frequently updated, the satellites accessed by each user within a given time slice remain fixed. In this paper, the satellite serving the ground source is defined as the ‘source’, and the satellite pointing to the destination address in the last hop is defined as the ‘sink’. Several SDN control stations deployed on the ground plan the routing paths from each source to the sink based on the network-wide information, including the network topology, link status, and user access information. In order to achieve network hierarchy separation and flexible data path management, we introduce the concept of network tunneling, where SDN controllers encapsulate data in specific protocols for transmission through tunneling technology when planning routing paths. This approach avoids privacy issues when transmitting data across domains and simultaneously enables more granular traffic regulation.

Based on the proposed model, one can implement global and local routing strategies under large-scale low-orbit satellites. After the SDN controller completes the path computation, the computed routing information will be updated in the whole network through the flooding mechanism. Considering the suddenness of traffic and the dynamic changes of inter-satellite links, our scheme not only relies on the historical traffic data to optimally adjust the traffic of the whole network periodically, but also designs a robust rerouting mechanism to cope with the failure of local links or satellites in order to ensure the stability and reliability of the system.

\begin{figure}[t]
\centering
\includegraphics[scale=0.34]{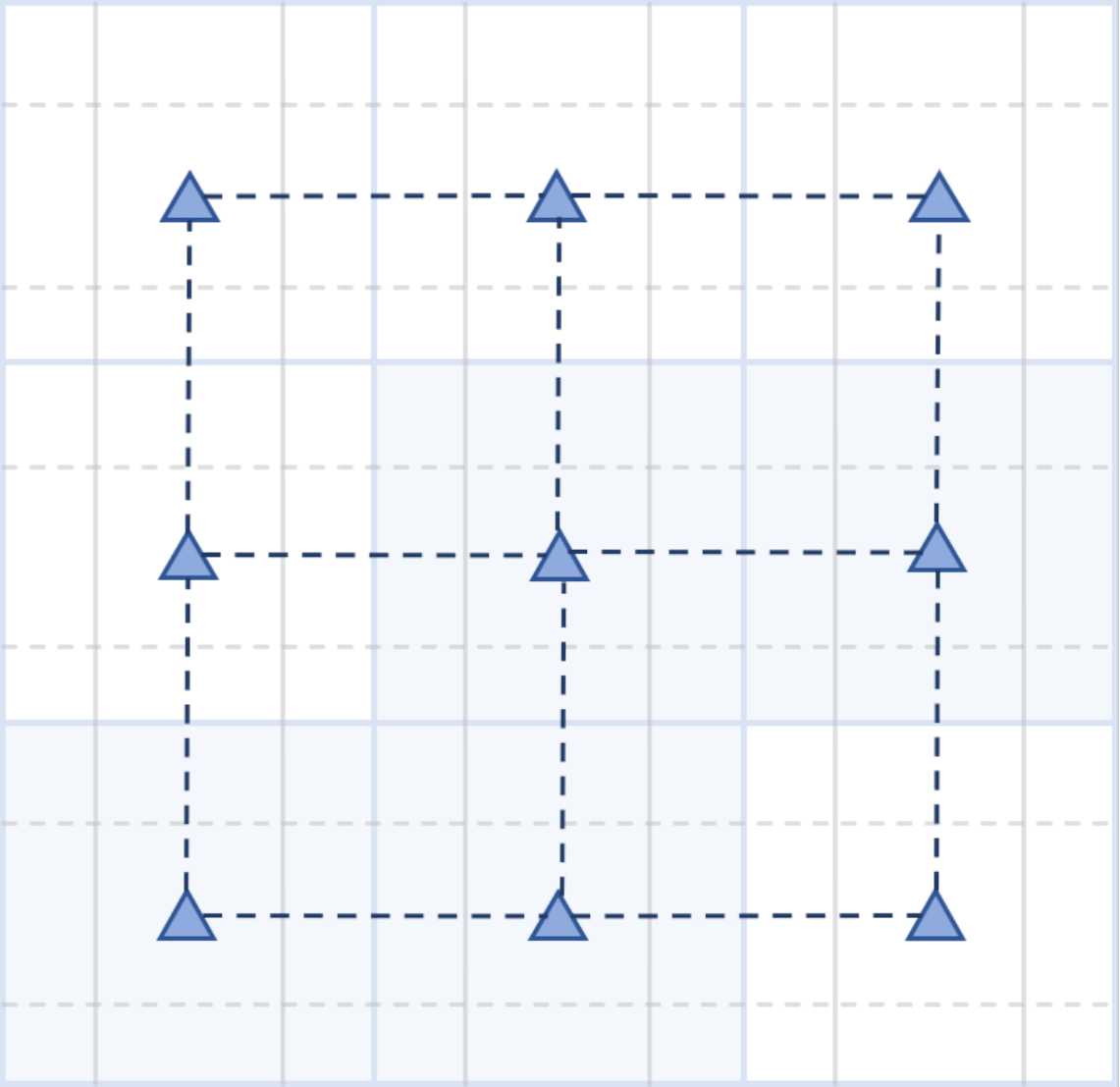}
\caption{Partition Schematic Diagram}
\label{partition}
\end{figure}

\section{Lightweight Design of Probabilistic Multipath Segmented Routing Based on Skeleton Path}
\subsection{Skeleton Routing Planning in Satellite Networks}
In large-scale satellite constellations, obtaining the complete forwarding path for each tunnel is challenging and time-consuming, making real-time processing infeasible. When planning forwarding paths for a large number of tunnels simultaneously, inter-satellite link congestion may occur, leading to uneven traffic distribution. To address this, we adopt a top-down, divide-and-conquer approach to partition the satellite network. As shown in Figure \ref{partition}, we divide the satellite constellation into \(m\) regions \(R\), each containing \(n \times n\) satellites. It's important to note that we can use irregularly shaped partitions to facilitate efficient real-time segmented planning of tunnel forwarding paths and ensure optimal network performance.  Square-based partitioning, in particular, makes landmark selection and skeleton graph acquisition easier. In our model, network boundaries in polar regions do not need to adhere to regular shapes or square regions. After partitioning, we assign a virtual landmark \(L\) to each partition. This landmark represents the direction of each tunnel between different regions, guiding the data flow to ensure it reaches the destination node. Typically, the landmark is set at the center of the square region, but it can be placed at any satellite node or position within the region, except for nodes on the region's boundaries.

We define the skeleton graph as a topological abstraction of the partition, representing the adjacency relationships of the partitions or landmarks to assist tunnel routing. The skeleton graph is represented as a directed graph \(G = (V, E)\), where vertices \(V\) represent regions or landmarks, and edges \(E\) represent the adjacency relationships between regions, referred to as skeleton links. These edges can be weighted to reflect the current network state characteristics, such as congestion or delay. Once constructed, the skeleton graph guides the routing of numerous tunnels and enables Traffic Engineering (TE). If we use square-based partitioning, the skeleton graph forms a regular grid topology. For irregular partitioning, we construct the skeleton graph using a Voronoi diagram.
\begin{figure}[t]
\centering
\includegraphics[scale=0.2]{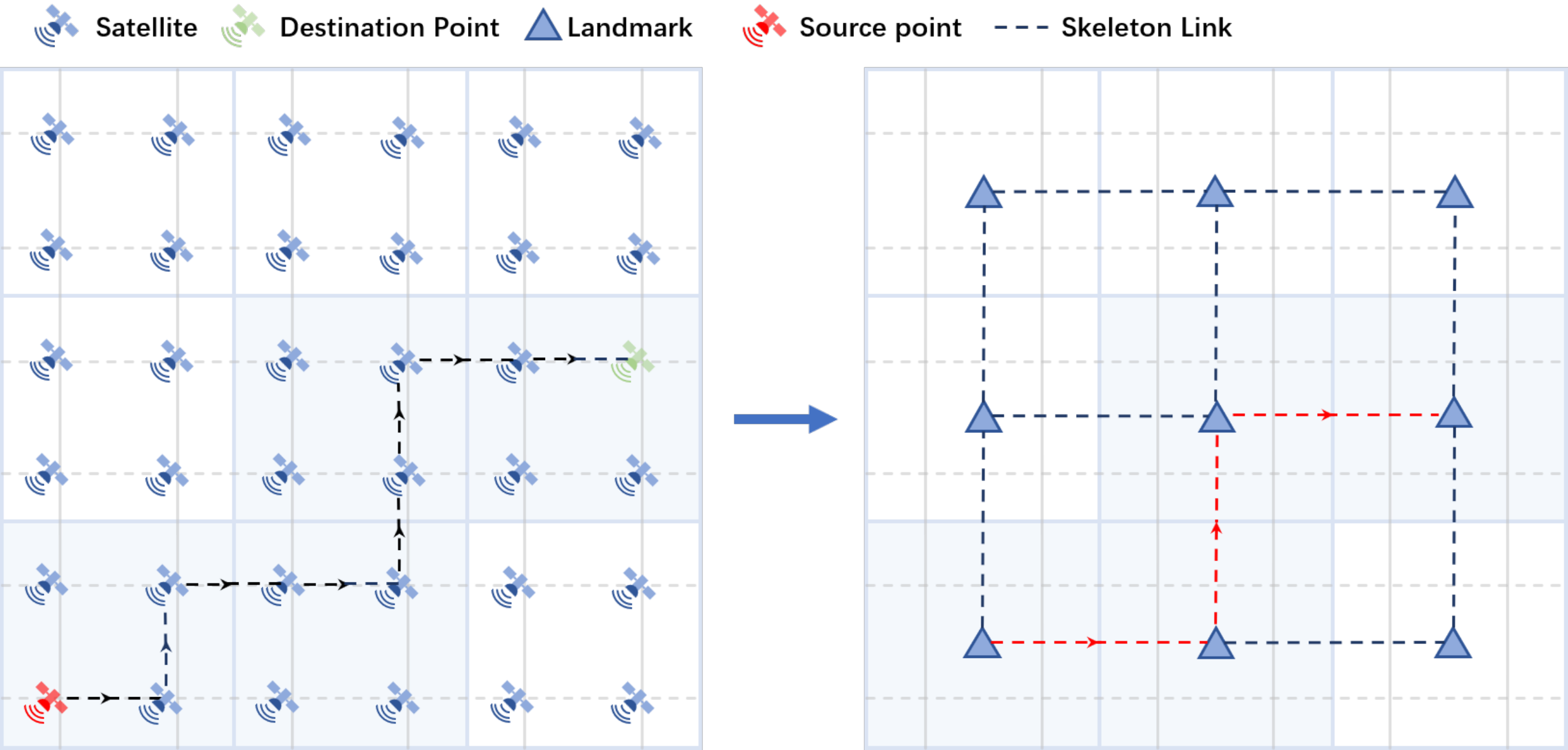}
\caption{Illustration of a Skeleton Path in the Network}
\label{skeleton_path}
\end{figure}
\subsection{Construction of Skeleton Path}
Once the skeleton graph is constructed using global network information, as shown in Figure \ref{skeleton_path}, the controller generates a skeleton path \( P \) for each tunnel according to the source and destination points. Each skeleton path \( P \) is a collection of landmarks \( L \). The skeleton path depicted in the figure is composed of the landmark set \(\{L_{0}, L_{1}, L_4, L_5\}\). This path sequentially traverses these critical landmark nodes, facilitating data transmission between the source and destination nodes. If multiple tunnels have their source and destination points within the same regions and share the same quality of service classification, they can use the same skeleton path without reconstructing it.

To ensure the efficiency and optimality of the skeleton path, we propose the Landmark-based Geometric Segment Routing (LGSR) algorithm. This algorithm generates skeleton paths using weighted parallel processing, balancing traffic load on the skeleton links while ensuring the shortest paths, and then distributes these routes to the respective source nodes. Essentially, a skeleton path \( p_{m,n} \) represents the inter-region routing from the source region \( R_m \) to the destination region \( R_n \). Since all vertices in the skeleton graph are landmarks, the skeleton path indicates which regions the tunnel should traverse, providing a coarse-grained guide for traffic forwarding.

The LGSR algorithm proceeds as follows: First, the skeleton path weights \( W \) are initialized, and the satellite set \( S \) is partitioned into multiple regions \( R \). Based on these regions \( R \) and associated landmarks \( L \), the skeleton graph \( G \) is constructed. The algorithm then iterates over each flow in the flow tunnel set \( F \), where each flow tunnel consists of the source satellite \( S_i \), destination satellite \( S_j \), packet size \( F \), and an offset parameter \( offset \). Using the source satellite \( S_i \) and destination satellite \( S_j \), the corresponding landmarks \( L_m \) and \( L_n \) within their respective regions are determined. The Dijkstra algorithm is subsequently applied to compute the skeleton path \( P \), which represents the path from landmark \( L_m \) to \( L_n \).
After obtaining the skeleton path \( P \), the algorithm computes the rectangular band-shaped region between adjacent landmarks along the path, and traffic is broadcast within these regions. Simultaneously, the flow distribution is adjusted based on the offset parameter \( offset \). After a specified amount of traffic has been processed (determined by the parameter \( batch \)), the algorithm retrieves the traffic status of the current region and updates the skeleton path weights \( W \) according to the observed traffic conditions. Finally, the computed skeleton paths \( P \) for each tunnel are returned.

\begin{algorithm}[H]
  	\caption{LGSR Algorithm}
  	\label{alg:LGSR}
  	\begin{algorithmic}[1]
  		\Require Satellites: $S$, Flow Tunnels: $T=\{S_i, S_j, F, offset\}$, Update batch: $batch$, Update parameter: $\alpha$
  		\Ensure Skeleton Path $P$
  		\State Initialize skeleton path weights $W$
  		\State Divide satellites \(S\) into regions \(R\)
            \State Initialize skeleton links \(E\)
  		\State Construct skeleton graph: $G = \{R, L, E, W\}$
  		\For{$tunnel\_id, flow\_tunnel$ in $T$}
            \State Get current node \(v=(x_v,y_v)\)
  		\State Retrieve landmarks \(L_m\) and \(L_n\) based on source satell-
        \Statex \quad  ite \(S_i\) and destination satellite \(S_j\) in the flow
            \State $P = Dijkstra(L_m, L_n, G)$
            \State Retrieve the $offset$ value from $flow\_tunnel$
            \For{$L_{i+1}$ in $P$}
            \State \(v=\)\textbf{Adaptive\_Forwarding}(\(G\), \(v\), \(L_{i+1}\),  \(offset\) ) 
            \EndFor
            \If {$tunnel\_id \% batch == 0$}
                \State \textbf{Update\_skeleton\_weights($G$, $\alpha$)} to update link 
                \Statex \quad \quad \quad weights
  		\EndIf
  		\EndFor
            \State \Return $P$
  	\end{algorithmic}
\end{algorithm}

The skeleton path serves as a sequence of landmarks guiding the actual routing of data packets. To implement inter-region routing, we construct a landmark-based segment list for segmented routing between regions. During the data packet forwarding process from the source node to the destination node, each region \( R_m \) and \( R_n \) corresponds to a segment, i.e., a skeleton link \( P_{m,n} \). An SDN-based segmented routing strategy allows for global information awareness of the satellite constellation, ensuring each tunnel follows the shortest path with balanced traffic, preventing skeleton link congestion.

Specifically, during the initialization of the packet header at the source node, the controller generates the segmented list information based on the skeleton path and sends it to the source nodes of each tunnel. This information is stored in the routing table of the source node. For outgoing packets, the source node checks the routing table to obtain the corresponding segment list, processing it step by step to ensure routing efficiency and minimal satellite storage consumption. The relevant fields are written into the packet header, and the packet is forwarded towards the next landmark.

\begin{algorithm}[H]
    \caption{Update Skeleton Weights Algorithm}
    \label{alg:updateSkeletonWeights}
    \begin{algorithmic}[2]
    \Require Skeleton graph: $G$, Update parameter: $\alpha$
    \State Retrieve the current region's traffic: $TS = \{traffic, E\}$
    \State Initialize the traffic dictionary \(traffic\_dict\)
    \For{$traffic$, $E$ in $TS$}
    \State $traffic\_dict[E] += traffic$
    \EndFor
    \State $total\_traffic = \sum{traffic\_dict}$
    \For{$E, W$ in $G$}
    \State $norm\_traffic = traffic\_dict[E] / total\_traffic$
    \State $updated\_W = \alpha \cdot norm\_traffic + (1-\alpha) \cdot W$
    \State $G[E].W = updated\_W$
    \EndFor
  \end{algorithmic}
\end{algorithm}

\subsection{Probabilistic Forwarding of Data Packets}

Given the relatively small scale of satellite networks within each region, we designed a routing algorithm based on probabilistic forwarding to ensure balanced traffic distribution. Specifically, each satellite node uses deterministic forwarding probabilities for inter-satellite links in four different directions, ensuring data packets are always forwarded towards the next landmark or destination.

\begin{figure}[t]
\centering
\includegraphics[scale=0.32]{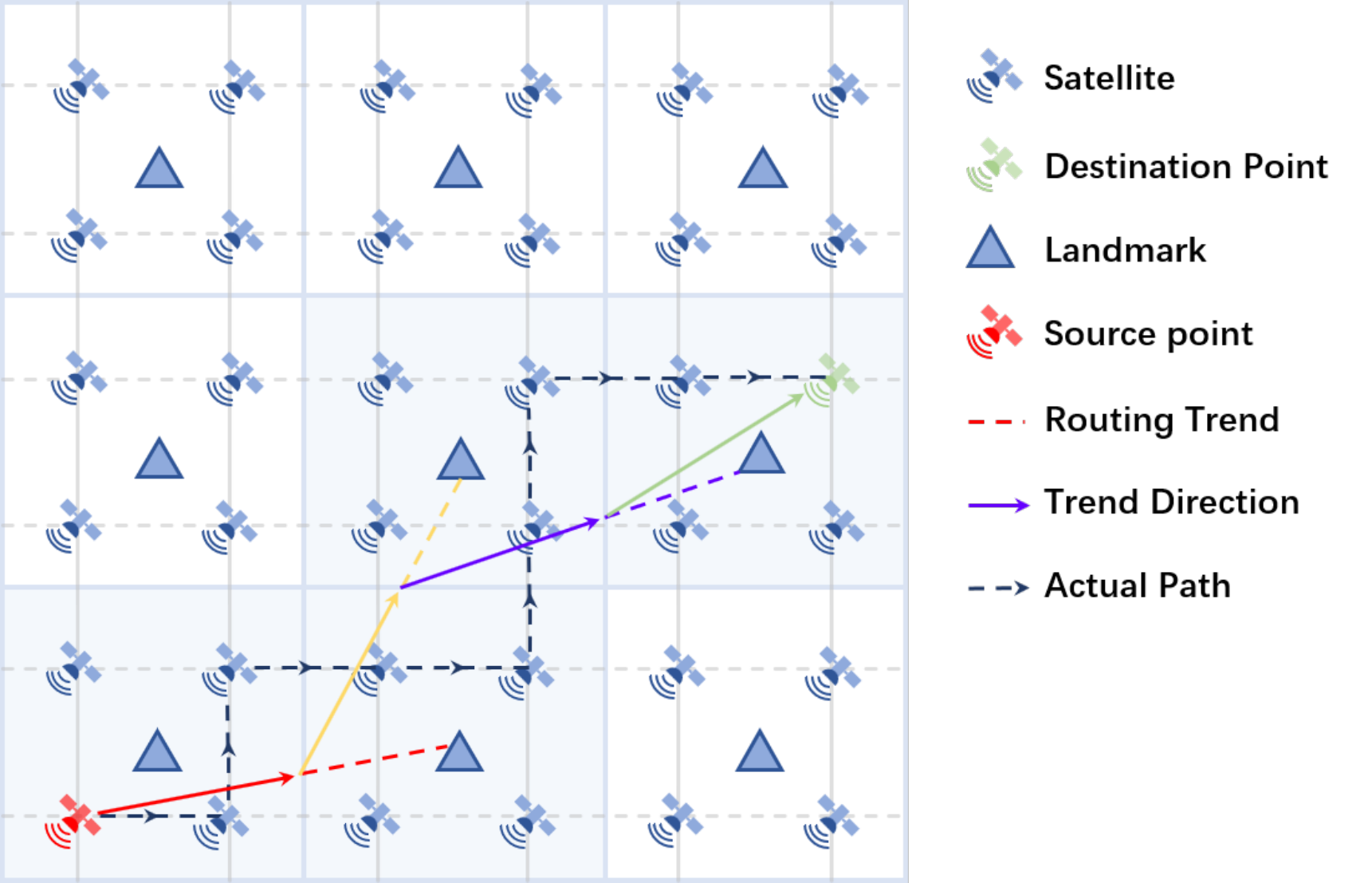}
\caption{Probabilistic Forwarding within Skeleton Path}
\label{qushitu}
\end{figure}

First, each satellite node determines the relative direction of the next landmark. Based on the shortest path rule, as shown in Figure \ref{qushitu}, a satellite node in a skeleton path will only perform probabilistic forwarding in two directions towards the destination. Each node knows the landmark of its region, so it checks the packet header’s label to find the next region's landmark in the landmark (segment) list field. If the current landmark is the final one, the packet will be forwarded directly towards the destination.

Under this strategy, within a region, before reaching the next region's boundary, the data packet is forwarded towards the next landmark. Upon reaching the boundary of the next region, the landmark is updated and the data packet is forwarded in the direction of the new landmark.

\begin{figure}[t]
\centering
\includegraphics[scale=0.34]{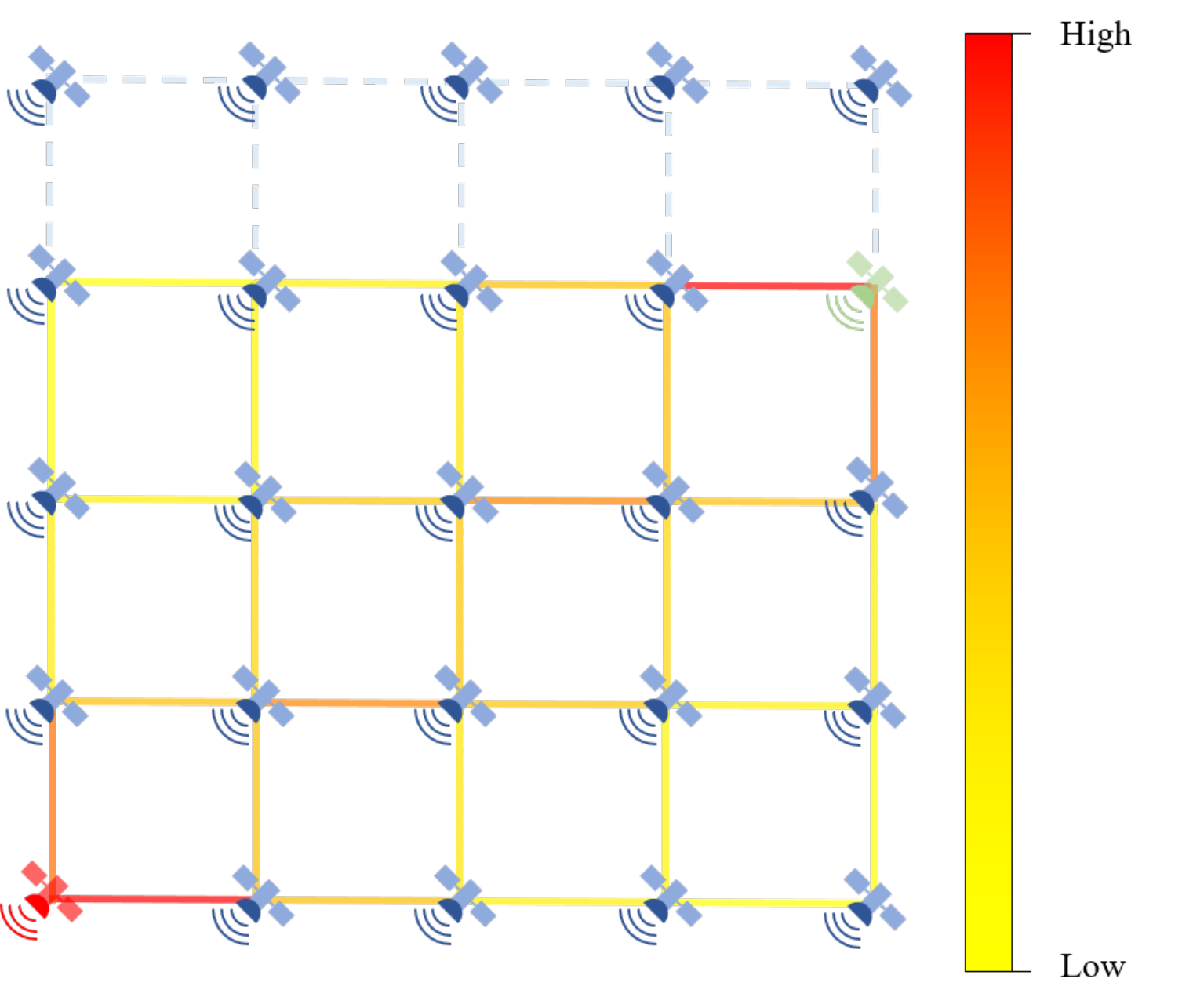}
\caption{Flow Distribution Visualization}
\label{xiaoguotu}
\end{figure}

Next, we determine the forwarding probability of each satellite towards different links. The implementation is as follows: since the node knows the coordinates of the target landmark (or destination) and all its neighbors, as well as its own coordinates, it can easily determine which neighbor to forward the packet to. By default, the packet can be forwarded with equal probability to the two neighbors in the direction of the landmark. These two neighbors are located within the rectangular area defined by the target landmark and the node itself, forming the diagonal vertices of the rectangle. If the target landmark and the node share the same \(x\) or \(y\) coordinate, the node simply forwards the packet to the neighbor in the direction of the landmark. The final result is shown in Figure \ref{xiaoguotu}.

\begin{algorithm}[H]
\caption{Adaptive Forwarding Algorithm}
\label{alg:probabilistic_forwarding}
\begin{algorithmic}[1]
\Require Skeleton graph $G$, Next-hop landmarks $L_{i+1}$, Offset $(\delta_x, \delta_y)$, Current node $v=(x_v, y_v)$
\Ensure Balanced traffic forwarding between $v$ and $L_{i+1}$, Updated destination node \(v\)
\State Compute the initial rectangular region \[\text{Rect} = (x_v, y_v) \to (x_{L_{i+1}}, y_{L_{i+1}})\]
\State Apply offset $(\delta_x, \delta_y)$ to $\text{Rect}$
\[
\text{Rect}_{\text{adjusted}} = (x_{v}, y_{v})  \to  (x_{L_{i+1}} + \delta_x, y_{L_{i+1}} + \delta_y)
\]
\If{$v$ shares the same $x$ or $y$ coordinate with $L_{i+1}$}
    \State Forward packet deterministically towards $L_{i+1}$
\Else
    \State Determine two neighbors $n_1$ and $n_2$ within $\text{Rect}_{\text{adjusted}}$
    \State Calculate the relative probabilities for $n_1$ and $n_2$ based on:
    \[
    P(n_1) = \frac{|x_{L_{i+1}} - x_{v}| }{|x_{L_{i+1}} - x_v| + |y_{L_{i+1}} - y_v|}
    \]
    \[
    P(n_2) = 1 - P(n_1)
    \]
    \State Forward packet probabilistically to $n_1$ or $n_2$ until \(v\) reaches the boundary
\EndIf
\State Update the cumulative traffic distribution matrix $T$
\State \Return Updated destination node $v$
\end{algorithmic}
\end{algorithm}

However, due to the cumulative effect of probabilistic forwarding, the links near the target node often experience a significant increase in traffic load, leading to highly concentrated traffic around the target node and resulting in load pressure. To alleviate this issue, the algorithm introduces an offset tuple \( offset = (\delta_x, \delta_y) \) to dynamically adjust the boundaries of the band-shaped region through random offsetting. The range of the band-shaped region is determined by the region boundaries, the starting landmark, and the target landmark. When calculating the band-shaped region, a random offset \( offset \) is applied to dynamically adjust the region boundaries, where \( \delta_x \) and \( \delta_y \) can be positive or negative values. These offsets adjust the region boundaries by shifting them upward by \( \delta_x \) units or leftward by \( \delta_y \) units.
This offset directly affects the coordinates of the target landmark. The original coordinates of the landmark \((x_{L_i}, y_{L_i})\) are adjusted to \((x_{L_i} + \delta_x, y_{L_i} + \delta_y)\). Consequently, the original range of the band-shaped region \((x_v, y_v) \to (x_{L_{i+1}}, y_{L_{i+1}})\) is updated to \((x_v, y_v) \to (x_{L_{i+1}} + \delta_x, y_{L_{i+1}} + \delta_y)\). The detailed implementation process is outlined in Algorithm \ref{alg:probabilistic_forwarding}.
Through this dynamic adjustment mechanism, the algorithm optimizes traffic distribution within the band-shaped region, effectively alleviating traffic concentration caused by probabilistic forwarding, thereby enhancing link load balancing and improving the overall network performance.

This routing algorithm inherently provides load balancing. Compared to single-path routing, this probabilistic multipath routing scheme distributes the traffic of a tunnel more evenly over a broader area from the source to the destination. This area is called a flow domain, and the larger the flow domain, the wider the traffic distribution range, leading to less traffic on each link and better load balancing. It is TE-friendly because the traffic of each tunnel is spread over a larger flow domain area, reducing the impact of individual tunnel optimizations on other tunnels' performance. Through this probabilistic multipath forwarding method, we can ensure balanced traffic distribution on each link within the region, avoiding inter-satellite link congestion without requiring complex routing algorithms. This approach ensures efficient and real-time performance of intra-region traffic forwarding.

\section{Hierarchical Segment Routing: Local Segment Routing Based on Local Landmarks}
\subsection{Local Skeleton Graph and Hierarchical Routing Planning}
\begin{figure*}[t]
\centering
\includegraphics[scale=0.36]{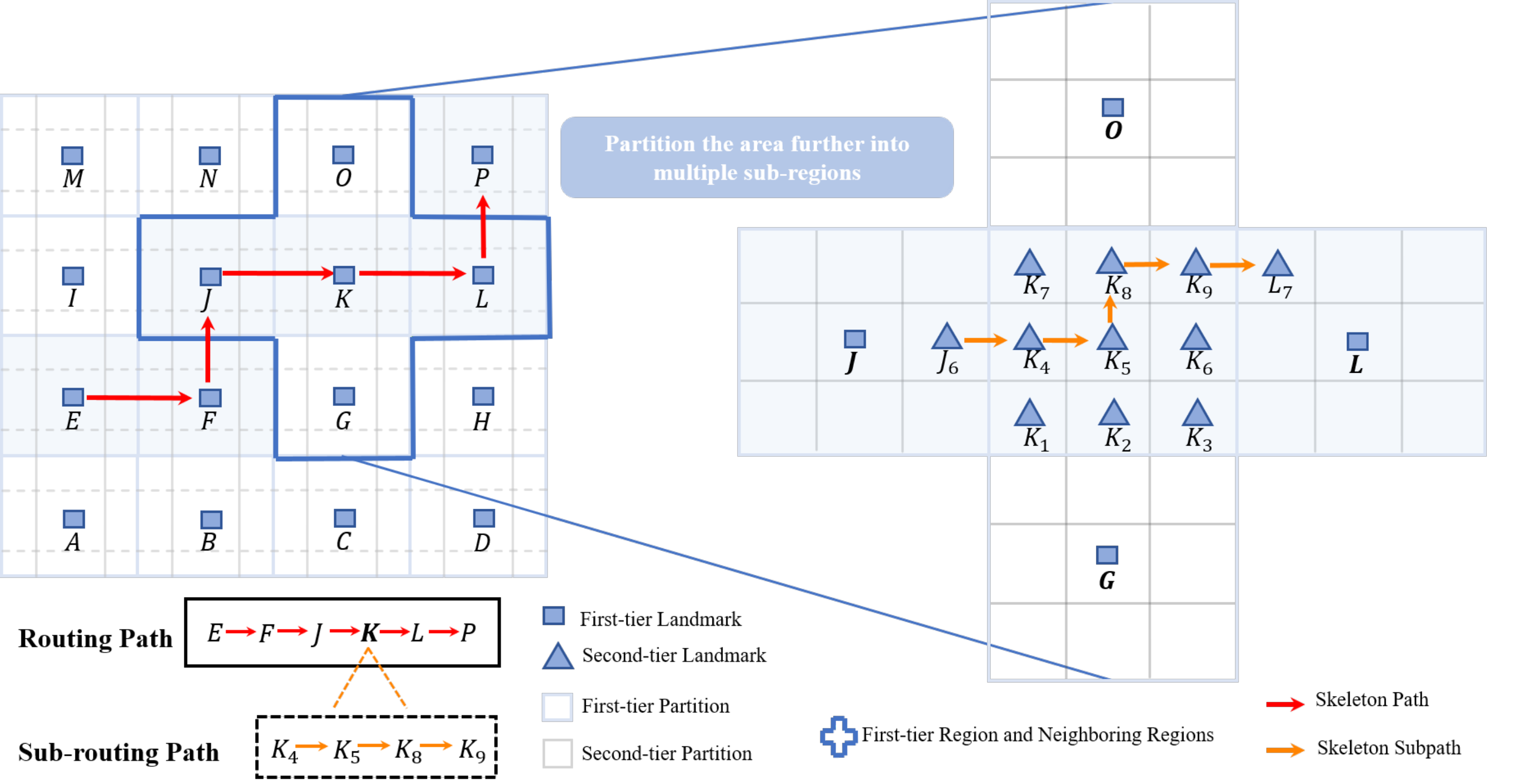}
\caption{Regional Segmented Routing Design}
\label{Regional Segmented Routing Design}
\end{figure*}

Given the highly dynamic nature of space networks, even with routing planning within individual regions, a high volume of concurrent tunnels can still result in service congestion and quality degradation. To address this, hierarchical routing planning within regions becomes essential. As illustrated in Figure \ref{Regional Segmented Routing Design}, each large region can be subdivided into smaller sub-regions, each containing local landmarks, thereby forming a multi-layered landmark structure and enabling hierarchical partitioned routing. In the figure, the region containing landmark K is partitioned into several sub-regions \(\{K_1, K_2, K_3, \dots, K_9\}\). A local skeleton graph is then constructed for this region, with edge weights assigned in a manner consistent with the global skeleton graph. Subsequently, a greedy routing algorithm is applied to derive the local skeleton path, which is used to guide local source routing. Ultimately, this approach yields a more granular skeleton sub-path \(\{J_6, K_4, K_5, K_8, K_9, L_7\}\), ensuring more efficient traffic management and routing within the region.

To implement local source routing from a node to its neighboring landmarks outside the sub-region, we need to construct an extended local skeleton graph. This local skeleton graph includes the sub-region itself and all neighboring landmark regions, updating the edge weights based on network conditions (e.g., congestion and delay). After obtaining the local skeleton graph and edge weights, we plan local skeleton paths and assign them to the corresponding local sources. The local skeleton path includes a list of local landmarks indicating the routing direction within the sub-region or to the next local landmark. When service congestion or quality degradation occurs, multi-layer landmarks for a given sub-region are constructed, and hierarchical partitioning routing is activated. Hierarchical partitioning routing is a top-down, divide-and-conquer approach that decomposes large-scale satellite constellations into smaller, manageable satellite groups for traffic optimization. Compared to single-layer partitioning, this refined hierarchical partitioning alleviates the complex pressure of numerous tunnels in ultra-large-scale satellite constellations, breaking down coarse-grained skeleton paths into multiple local skeleton graphs. Each skeleton link contains several local skeleton graphs, distributing tunnel traffic efficiently and evenly across each sub-region. Consequently, controllers can ensure the shortest path and traffic optimization through online real-time management and scheduling.

For instance, for a constellation with 4000 nodes, at most two layers of partitioning are needed. Assuming the basic block unit is \(2\times2\), the first-level (global) partitioning comprises \(6\times6=36\) blocks. Each first-level partition includes \(3\times3\) second-level (local) partitions, requiring at most 120 first-level partitions and up to 1080 second-level partitions. Another example: with a basic unit of \(4\times4\), the global partitioning consists of \(12\times12\) blocks, with each global partition containing \(3\times3\) local partitions, requiring 32 global partitions and \(288\) local partitions.

In this scenario (multi-layer local sub-regions), intermediate satellite nodes may become local sources with a certain probability, requiring local source routing to the next neighboring sub-landmark. Based on the local skeleton path, these local source nodes add new routing information to the packet header, and the packet is forwarded along the local skeleton path to the next local landmark. If a node needs to be disabled or restricted, weights and routes are recalculated within the local sub-region, avoiding global control and resource waste. Although the extended local skeleton graph includes the local sub-region and adjacent sub-landmark regions, the local skeleton path only includes local landmarks within the sub-region. Efficient local skeleton paths to neighboring landmarks can still be obtained. If the local aggregation point is the current target landmark, before reaching the regional boundary, traffic packets are forwarded along the local skeleton path to the next local landmark. Once the packet reaches the regional boundary of the target landmark, the local source routing segment terminates, and all relevant landmarks are updated accordingly. Therefore, local packet forwarding always occurs within the sub-region. This approach enables the network to better handle service congestion and quality degradation in highly dynamic environments, thereby improving overall network performance.

When encountering millions of concurrent tunnels, the hierarchical management and planning of hierarchical partitioning routing offer good scalability and flexibility. If link load balancing is achieved, hierarchical partitioning routing is not needed. If certain first-level partitions face performance bottlenecks or load pressure, more sub-landmarks can be flexibly marked to bypass congested areas, ensuring overall network stability and high performance.
\subsection{Hierarchical Routing Strategy with Local Table Routing and Local Source Routing}

Due to the rapid movement of Low Earth Orbit (LEO) satellites and drastic fluctuations in network traffic, certain satellites and their inter-satellite links may become congested or unavailable. These unavailable links can cause probabilistic multipath routing to fail, resulting in link paralysis. Therefore, enabling local table routing within the affected areas is necessary to replace probabilistic multipath routing and mitigate fluctuations, failures, and congestion.

The specific solution is as follows: when link failure information is timely reported to the controller, the local routing information of the affected area is updated in real-time. During the activation of local table routing, probabilistic multipath routing across regions is disabled, and nodes within the affected area use deterministic local routing (i.e., precise path forwarding for each tunnel). If the tunnel's target node is within the current area, the current routing segment is the final segment, and the shortest path algorithm can efficiently generate the routing path based on the current area's topology.

If the target node is outside the current area, the routing will find an appropriate path from the current node to the next landmark in the neighboring area. Notably, when traffic packets reach the boundary of the sub-region, the local routing planning stops, and the packets enter a new sub-region where boundary nodes replan the route to the next target landmark. Based on the divide-and-conquer approach, precise local routing planning is only conducted within sub-regions, ensuring the algorithm's online performance. For cross-region routing planning, a local map encompassing the current and next regions is used.

Thus, this local hierarchical routing planning supports both source routing and traditional table routing modes. When local source routing is enabled, the routing path can be stored in the boundary nodes of the region. When tunnel packets reach the boundary nodes, these nodes become the local sources for these packets. The boundary nodes search their routing table for the tunnel's local route, write this routing information into the packet header, and then forward the packet to the next target node.

When local table routing is enabled, the routing path is not stored in the boundary nodes. Instead, the tunnel's routing path is stored in all intermediate nodes along the path. Each intermediate node only stores the next-hop information of the tunnel in its local routing table and does not write the next-hop information into the packet hemader. When tunnel packets arrive at an intermediate node, the node searches its routing table for the next target node and forwards the packet to that node.

Combining probabilistic routing with local table routing offers strong adaptability, resistance to fluctuations, failures, and congestion, and is lightweight in terms of time, space, and complexity for both ground centers and satellites. It is suitable for rapid deployment and updates, and offers high flexibility with the ability to switch segment policies. When certain local areas experience failures or unavailable links, robust local routing within these areas can be activated.
\section{Proof}

\subsection{Analysis of Path Calculation and Update Transmission Efficiency}
Compared to using the Dijkstra algorithm directly, the LGSR method significantly reduces the path calculation and update transmission cost, as analyzed below.

\subsubsection{Path Calculation Cost}
Using LGSR to calculate tunnel paths reduces the cost by a factor of \(R^4\), meaning the calculation speed is improved by \(R^4\), where \(R\) is the number of satellites per region. This reduction is mainly due to the following two factors:

\paragraph{Algorithm complexity reduced by \(R^2\)} The complexity of the Dijkstra algorithm is \(N^2\), where \(N\) is the number of nodes in the satellite network. When applying the LGSR algorithm, the path calculation for each tunnel only requires the generation of the skeleton path. Since each region contains \(R\) satellites, the number of partitions is \(N / R\). Therefore, the size of the skeleton graph is \(N / R\), and the complexity of generating the skeleton path using the Dijkstra algorithm is \((N / R)^2\). Thus, LGSR reduces the computational complexity by a factor of \(R^2\) compared to using the Dijkstra algorithm directly.

\paragraph{Tunnel path count reduced by \(R^2\)} LGSR uses a tunnel aggregation technique, where tunnels with the same start and end points in the same region can share the same skeleton path. There can be up to \(R^2\) tunnels in each region, and they do not need to compute individual paths, thus reducing the number of tunnel paths to be calculated by a factor of \(R^2\). This tunnel aggregation technique significantly improves path calculation efficiency, especially as the region size increases, leading to greater performance improvements.

\subsubsection{Path Update Transmission Cost}
The path update transmission cost is also greatly reduced by LGSR. Compared to the Dijkstra algorithm, LGSR reduces transmission overhead by a factor of \(R^2 \sqrt{R}\). This reduction is due to the following reasons:

\paragraph{Skeleton path shortened by \(\sqrt{R}\)} Assuming the straight-line distance between the tunnel's start and end points is \(d\), the shortest path calculated by the Dijkstra algorithm would also be \(d\). For a square region with a side length of \(\sqrt{R}\), the skeleton path length for LGSR is approximately \(d / \sqrt{R}\), meaning the number of regions traversed is reduced by a factor of \(\sqrt{R}\), which reduces the communication transmission volume.

\paragraph{Tunnel path sharing reduces the update requirement by \(R^2\)} Since the skeleton path is shared, LGSR reduces the number of tunnel path updates by \(R^2\). This further reduces the amount of data transmission required and improves the efficiency of path updates.

By optimizing both path calculation and updates, LGSR provides significant improvements in computational complexity and transmission efficiency compared to the traditional Dijkstra algorithm, making it particularly suitable for large-scale satellite network environments.
\subsection{Shortest Path Analysis}  
Consider a grid network divided into several identical rectangular regions, forming a grid topology. For a Low Earth Orbit (LEO) satellite network, the skeleton map can also be easily represented as a grid topology. We aim to prove the following theorem: If the hop count \( H \) of the skeleton path equals the Manhattan distance \( D \) between the source landmark \( S \) and the destination landmark \( T \) on the skeleton map, plus one (i.e., \( H = D + 1 \)), then the actual forwarding path will follow the optimal path in terms of Inter-Satellite Link (ISL) hop count.

We assume that all landmarks of the tunnel are located within the rectangle determined by the source \( S \) and destination \( T \), i.e., the rectangle with \( S \) and \( T \) as diagonal vertices.

\begin{itemize}
    \item \( S = (x_S, y_S) \): Coordinates of the source landmark.
    \item \( T = (x_T, y_T) \): Coordinates of the destination landmark.
    \item \( \text{Rect}(S, T) \): The rectangle determined by \( S \) and \( T \).
    \item \( \text{Dist}(S, T) \): The Manhattan distance between \( S \) and \( T \), i.e., \( |x_T - x_S| + |y_T - y_S| \).
    \item \( H \): Hop count of the skeleton path.
    \item \( P \): The actual forwarding path.
\end{itemize}

The skeleton path is an idealized path, typically constructed by selecting a series of intermediate landmarks. We assume that the hop count \( H \) of the skeleton path satisfies the following relationship:
\begin{equation}
H = \text{Dist}(S, T) + 1
\end{equation}

The hop count \( H \) of the skeleton path is one more than the Manhattan distance \( \text{Dist}(S, T) \). This additional hop accounts for any potential extra overhead or deviation in the actual forwarding path, although this scenario typically does not occur if the skeleton path is already optimized.

Along the skeleton path, we construct several intermediate landmarks \( L_1, L_2, \ldots, L_n \), such that each subsequent landmark is located northeast of the previous one, i.e.,
\begin{equation}
\begin{split}
L_i = &(x_i, y_i), \quad x_S \leq x_1 \leq x_2 \leq \ldots \leq x_T, \\
&\quad y_S \leq y_1 \leq y_2 \leq \ldots \leq y_T
\end{split}
\end{equation}

The distance between each \( L_i \) and \( L_{i-1} \) is 1.

In each segment, the local destination landmark is northeast of the local source landmark. Suppose \( L_i \) is the source landmark of a segment and \( L_{i+1} \) is the destination landmark of that segment. The locally shortest path implies moving either north or east in each hop along the shortest path:
\begin{equation}
\text{Segment}_i = \{ (x, y) \mid x_i \leq x \leq x_{i+1}, y_i \leq y \leq y_{i+1} \}
\end{equation}
For each segment, the forwarding path is optimal because packets are forwarded along the shortest path within that segment.

By combining all segments, we observe that packets move along the shortest possible path at each hop, always progressing toward the northeast, until they reach the destination landmark \( T \). Therefore, the overall forwarding path is also the shortest possible path.
\subsection{Load Balancing Performance Analysis}
Theoretical derivation and experimental validation demonstrate that the LSGR algorithm significantly outperforms the Dijkstra algorithm in terms of load balancing. It effectively reduces unequal traffic distribution and prevents link overloading. Moreover, the LSGR algorithm disperses traffic, alleviates the load on hotspot links, and improves overall network performance and stability. 
\subsubsection{Average Load}
\begin{figure}[t]
\centering
\includegraphics[scale=0.43]{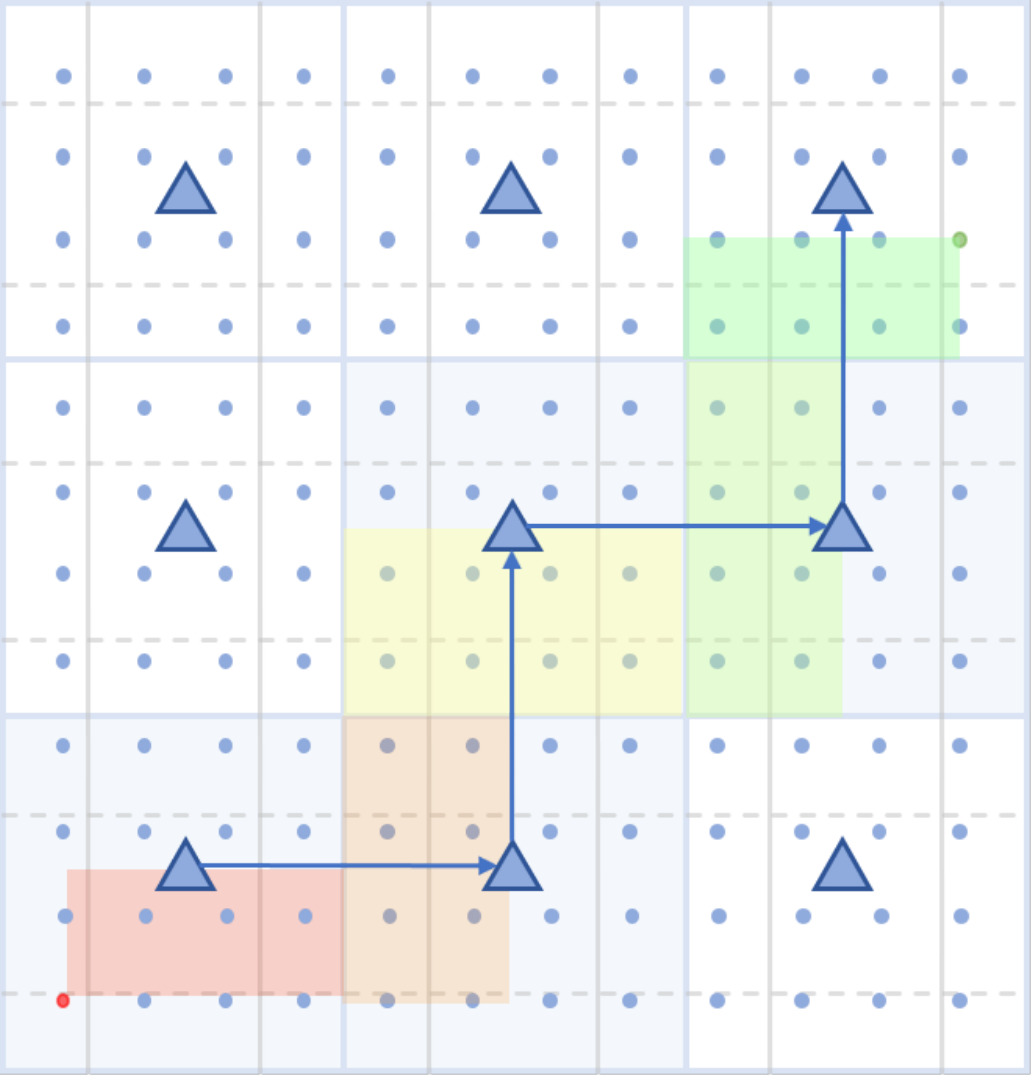}
\caption{Band-shaped Domain}
\label{band-shaped domain}
\end{figure}

Assuming the average width of the single tunnel domain (the region through which data packets are forwarded) is \(W\), the average traffic carried on each edge within the single tunnel domain for a single-path routing algorithm (e.g., Dijkstra) is \(1/W\), as illustrated in Figure \ref{band-shaped domain}.

\subsubsection{Upper Bound of Load on Links}
We consider a communication network composed of satellite nodes, which are interconnected via wireless links to facilitate information transmission and forwarding. In this network, skeleton routing can be easily derived based on the skeleton path and the known source and target satellites. As shown in the figure, probabilistic forwarding between the satellite source and the region boundary endpoint forms a probabilistic rectangular region between adjacent areas. Along the skeleton path, a series of elongated probabilistic rectangular regions are formed.

We define the width of one such region as \(a\) (representing the number of satellite nodes along the horizontal axis) and the length as \(b\) (representing the number of satellite nodes along the vertical axis). If the probabilities of a node forwarding traffic to the next landmark (or destination) through two ports are \(a/(a+b)\) and \(b/(a+b)\), respectively, and \(b > a\), then the upper bound of the traffic carried on each edge in this segment is \(b/(a+b)\). The proof is as follows.

Since \(b > a\), the maximum probability of forwarding traffic from the source node through the two ports is \(b/(a+b)\). As each node probabilistically forwards traffic through two ports, the probability on the next edge will be smaller than the current port's probability, continuing until the boundary of the region is reached. Therefore, the maximum forwarding probability in the triangular region formed by the boundary is the initial forwarding probability from the source node, \(b/(a+b)\). The maximum probability along the triangular region toward the destination can be expressed as:
\begin{equation}
\sum_{i=0}^{a-1} \binom{i+b-1}{i} \left(\frac{a}{a+b}\right)^i \left(\frac{b}{a+b}\right)^b
\end{equation}
Given that
\begin{equation}
\sum_{i=0}^{a-1} \binom{i+b-1}{i} \left(\frac{a}{a+b}\right)^i \left(\frac{b}{a+b}\right)^b < \frac{b}{a+b}
\end{equation}
always holds true, the upper bound of traffic carried on each edge is \(b/(a+b)\).

If the probabilities of forwarding traffic from a node to the next landmark (or destination) through two ports are \(a/(a+b)\) and \(b/(a+b)\), respectively, where \(b > a\), and both \(b\) and \(a\) are positive integers, then the upper bound of traffic carried on each edge in this segment is \(b/(a+b)\). The following inequalities are easily proved:
\begin{equation}
\sum_{i=0}^{a-1} \binom{i+b-1}{i} \left(\frac{a}{a+b}\right)^i \left(\frac{a}{a+b}\right)^b < \frac{b}{a+b}
\end{equation}
\begin{equation}
\sum_{j=0}^{b-1} \binom{a-1+j}{j} \left(\frac{a}{a+b}\right)^a \left(\frac{b}{a+b}\right)^j < \frac{b}{a+b}
\end{equation}
\section{Performance Evaluation}
In this chapter, we evaluate the performance of the proposed LGSR algorithm through a series of simulation experiments and conduct a comprehensive analysis of its performance in satellite networks of various scales. We simulate a Low Earth Orbit (LEO) satellite constellation with networks of varying sizes. Then compare the performance of the LGSR algorithm with several commonly used routing algorithms to validate its superiority in terms of load balancing and communication latency.
\subsection{Experimental Settings}
The experiments were conducted on a computer equipped with an Intel Core i5-1135G7 CPU @ 2.4 GHz and 16 GB of RAM. The experimental environment used Python as the programming language, with the NetworkX library for modeling and simulating the satellite network. The simulation process consisted of several steps: simulation parameter configuration, satellite network topology construction, communication traffic generation, and result analysis. The hop delay coefficient \(\alpha\) was set to 0.2, and the link delay coefficient \(\beta\) was set to 0.8.
To validate the effectiveness of the LGSR algorithm, the experiment selects three other routing algorithms as baseline comparisons. They are Dijkstra \cite{chen2023dijkstra}, Greedy \cite{lu2023online}, and RAN (random) \cite{fratty2023random}.
The satellite network was configured with six different scales: 16×20, 24×30, 32×40, 40×50, 48×60, and 56×70. The tunnel numbers were set to six levels: 2.5k, 5k, 10k, 20k, 40k, and 80k, and tunnel sizes were set to 0.75 Gbps, 1.5 Gbps, 3 Gbps, 6 Gbps, 12 Gbps, and 24 Gbps.
For experiments on varying network scales, the tunnel count was fixed at 20,000 and the tunnel size at 6 Gbps. For experiments on varying tunnel numbers, the network scale was fixed at 56×70 and tunnel size at 6 Gbps. In the experiments on varying tunnel sizes, the network scale was fixed at 48×60, with a tunnel count of 20,000.
\subsection{Metrics}
To evaluate the performance of satellite routing algorithms, three key metrics are used: Gini coefficient, latency, and execution time.

Gini Coefficient (\( Gini \)) measures load balancing by quantifying the degree of imbalance in link flow distribution. It is calculated as:

  \begin{equation}
  Gini =  \frac{\sum_{i=1}^{n-1}\sum_{j=i+1}^{n}|x_i-x_j|}{n^2 \cdot \frac{1}{n} \sum_{i=1}^{n} x_i}
  \end{equation}
where \( n \) represents the number of backbone links, and \( X = \{x_1, x_2, ..., x_n\} \) denotes the flow values on each link. A higher Gini coefficient indicates a poorer load balancing performance.

Latency (\( d_{total} \)) evaluates the average communication latency between nodes. It is defined as the weighted sum of hop delay and link delay:

  \begin{equation}
  d_{total} = \alpha \cdot d_{hop} + \beta \cdot \left( d_{link\_base} \cdot \exp\left( k \cdot \rho_{u,v} \right) \right)
  \end{equation}
where \( d_{hop} \) and \( d_{link\_base} \) denote base delays, and \( \rho_{u,v} \) represents link utilization, reflecting the impact of load on latency.

Execution Time (\( t \)) represents the time required for an algorithm to complete the path planning process, indicating its computational efficiency.
\begin{figure*}[htbp]
    \centering
    \begin{minipage}{0.23\linewidth}
        \centering
        \includegraphics[width=1\linewidth]{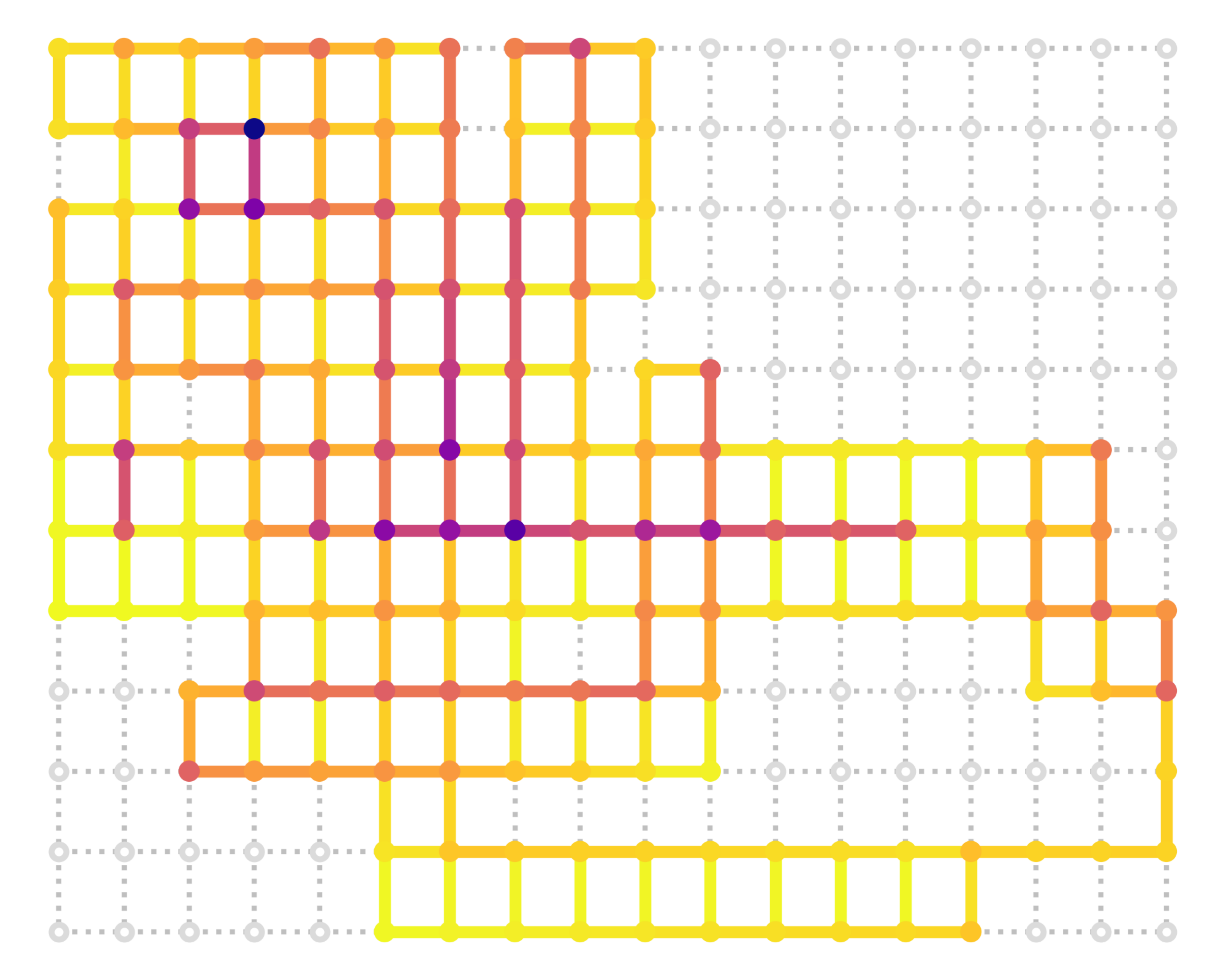}
        \caption*{(\textbf{a}) LGSR Algorithm}
        \label{fig:LGSR}
    \end{minipage}
    \hspace{0.01\linewidth}
    \begin{minipage}{0.23\linewidth}
        \centering
        \includegraphics[width=1\linewidth]{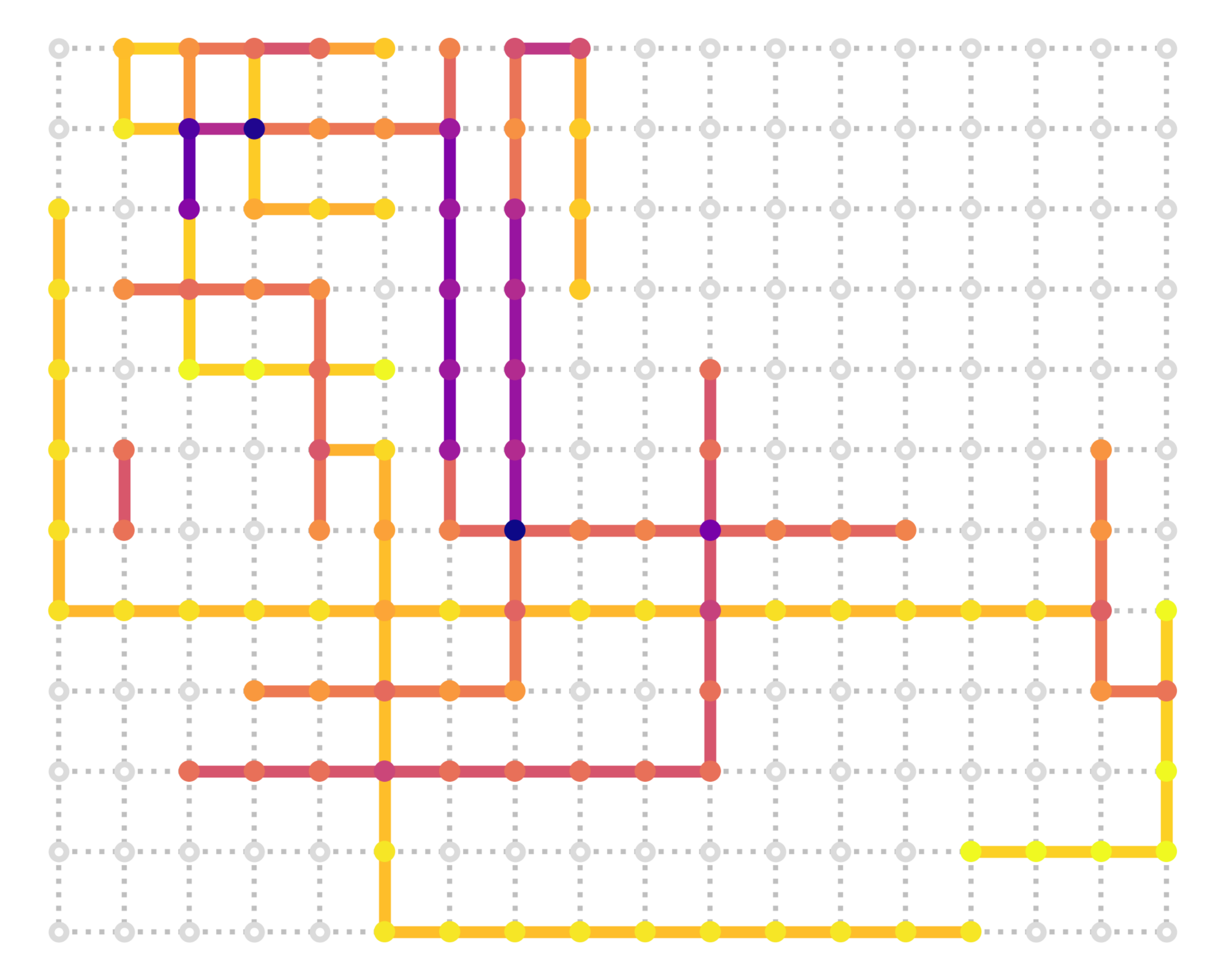}
        \caption*{(\textbf{b}) Dijkstra Algorithm}
        \label{fig:Dijkstra}
    \end{minipage}
    \vspace{0.5pt}
    \hspace{0.01\linewidth}
    \begin{minipage}{0.23\linewidth}
        \centering
        \includegraphics[width=1\linewidth]{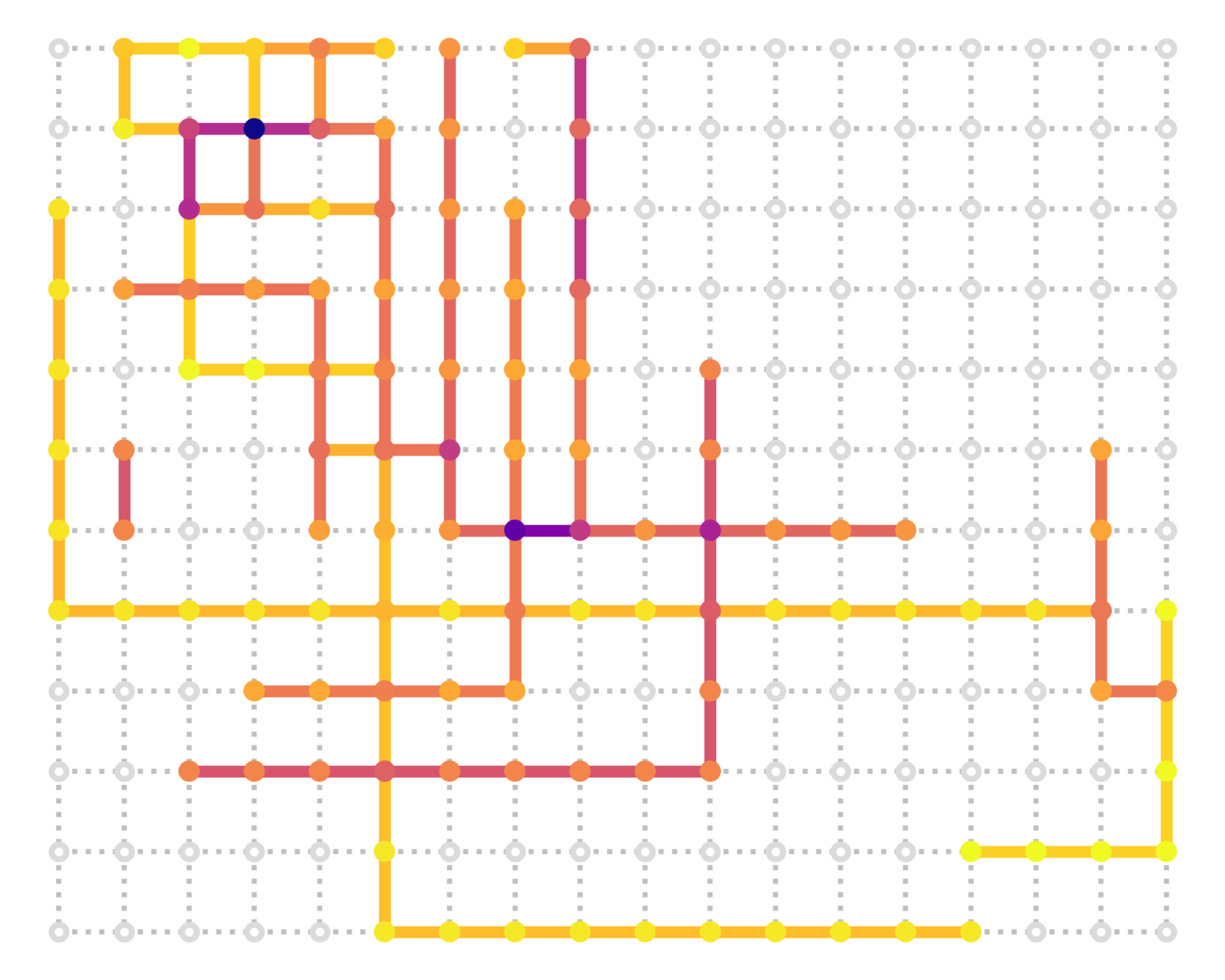}
        \caption*{(\textbf{c}) Greedy Algorithm}
        \label{fig:Greedy}
    \end{minipage}
    \hspace{0.01\linewidth}
    \begin{minipage}{0.23\linewidth}
        \centering
        \includegraphics[width=1\linewidth]{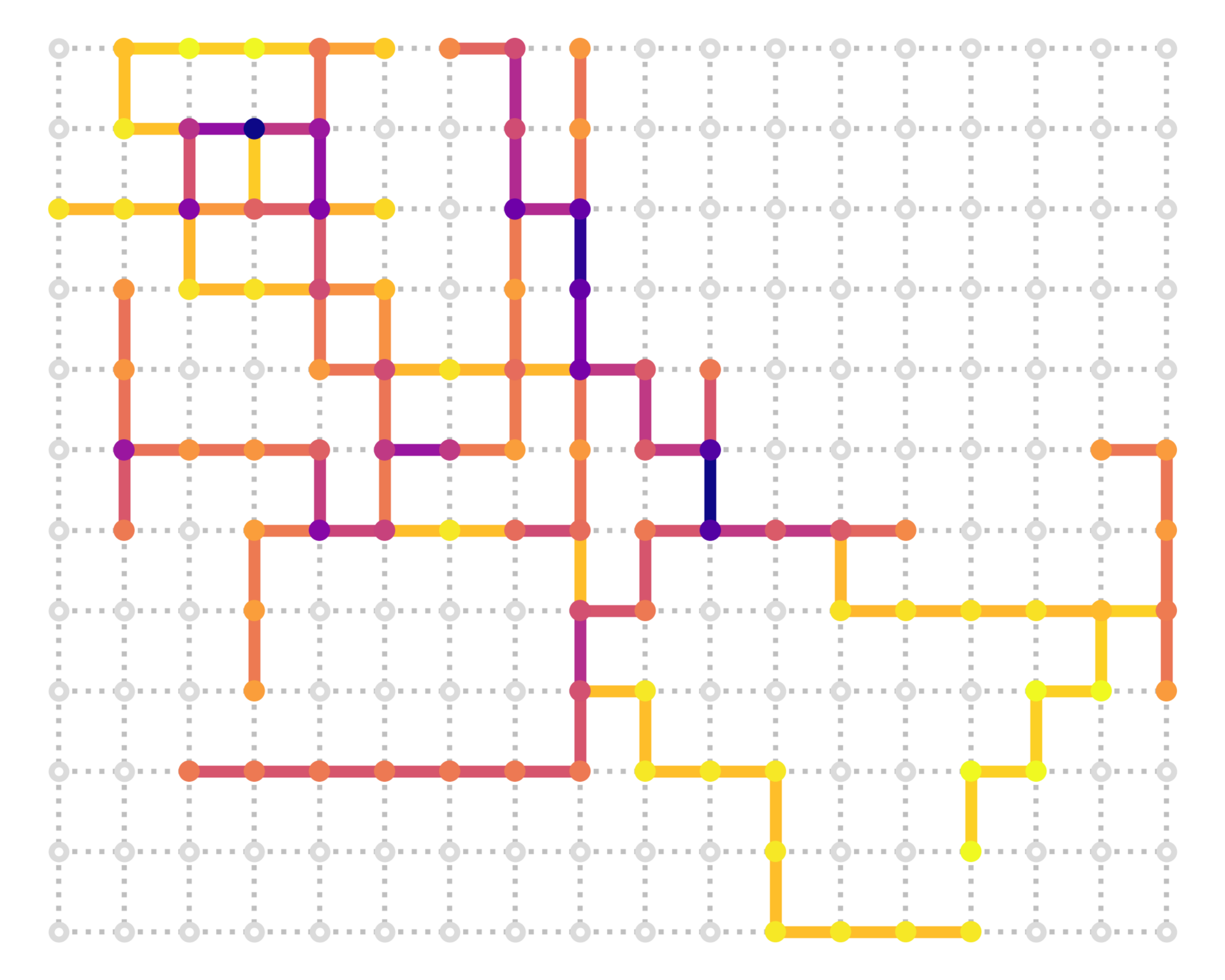}
        \caption*{(\textbf{d}) Random Algorithm}
        \label{fig:Random}
    \end{minipage}
    \caption{Path planning results of different algorithms in a 12×18 satellite network.}
    \label{fig:path_planning}
\end{figure*}

\begin{figure*}[htbp]
    \centering
    \begin{minipage}{0.31\linewidth}
        \centering
        \includegraphics[width=1\linewidth]{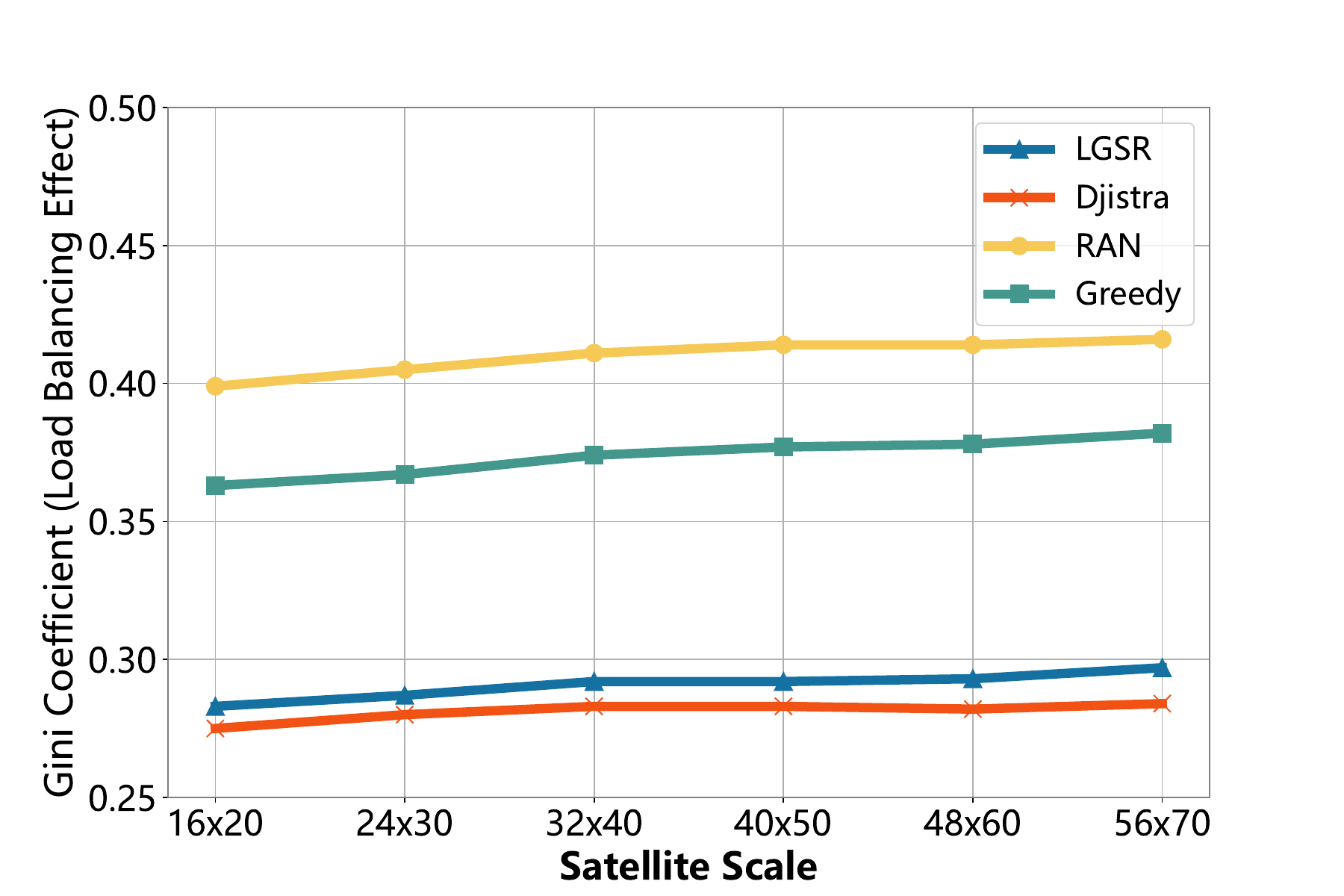}
        \caption*{(\textbf{a}) Load Balancing Performance by Satellite Scale}
        \label{fig:gini_scale}
    \end{minipage}
    \hspace{0.02\linewidth} 
    \begin{minipage}{0.31\linewidth}
        \centering
        \includegraphics[width=1\linewidth]{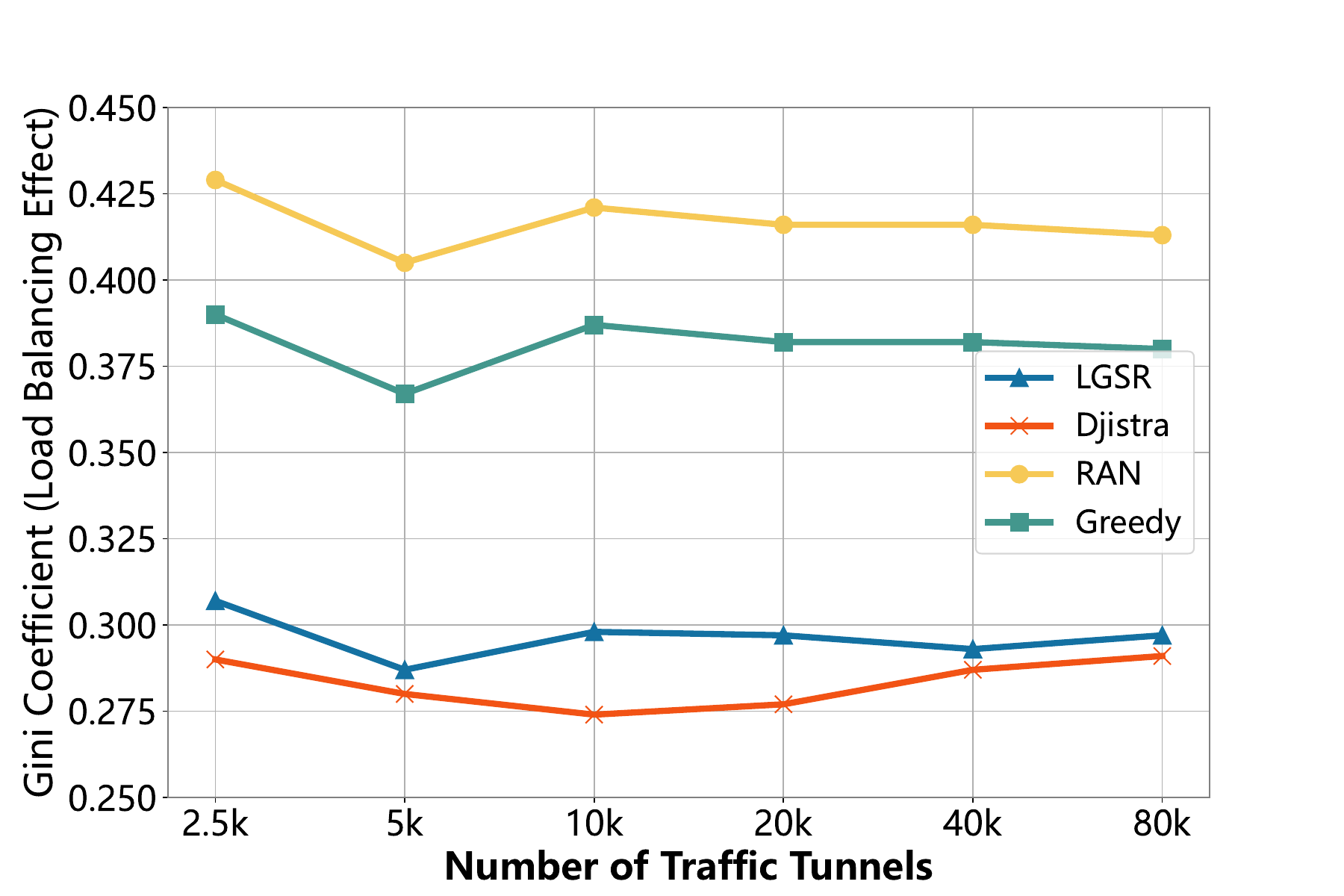}
        \caption*{(\textbf{b}) Load Balancing Performance by Number of Tunnels}
        \label{fig:gini_tunnels}
    \end{minipage}
    \hspace{0.02\linewidth} 
    \begin{minipage}{0.31\linewidth}
        \centering
        \includegraphics[width=1\linewidth]{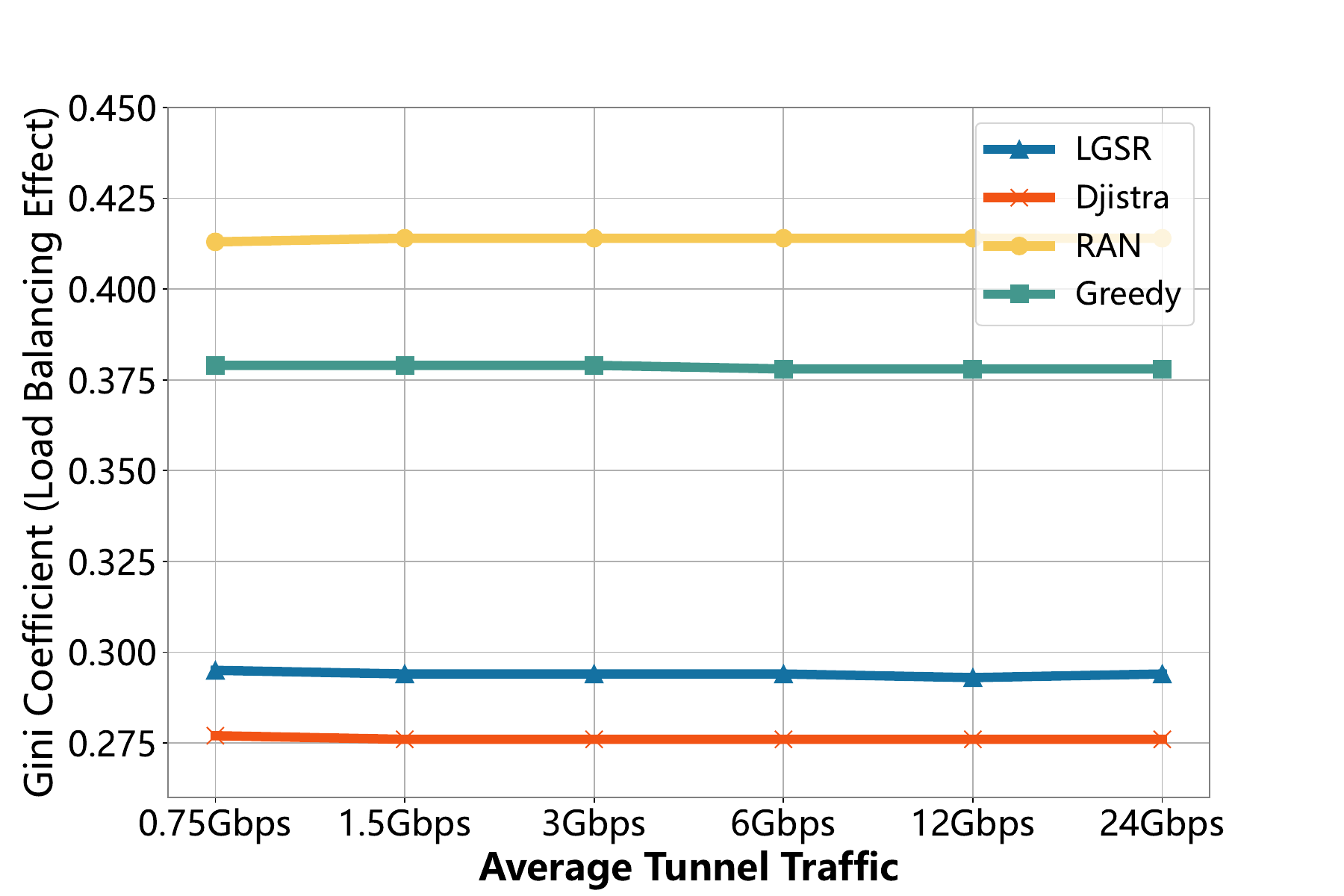}
        \caption*{(\textbf{c}) Load Balancing Performance by Tunnel Traffic Size}
        \label{fig:gini_avg_tunnel}
    \end{minipage}
    \caption{Load balancing performance under different conditions: (\textbf{a}) by satellite scale, (\textbf{b}) by number of tunnels, and (\textbf{c}) by tunnel traffic size.}
    \label{gini}
\end{figure*}

\subsection{Flow Distribution Analysis}
Figure \ref{fig:path_planning} shows the flow distribution of different algorithms under 20 traffic tunnels in a 12×18 satellite network. The results indicate that the LGSR algorithm achieves a more balanced flow distribution. Compared to other algorithms, the links in the LGSR distribution appear lighter in color, indicating lower load levels. Additionally, the presence of more branching paths suggests that LGSR is capable of dispersing traffic more evenly across the network, effectively reducing the risk of congestion on specific links.

In contrast, other algorithms rely on single-path routing strategies and fail to account for global load balancing, resulting in significant traffic accumulation on specific nodes and links. This is reflected by the darker link colors in the figure, which highlight the imbalance in load distribution and the increased likelihood of link congestion and communication bottlenecks, thereby affecting the overall network performance.
LGSR's ability to maintain balanced traffic distribution under various conditions is attributed to its probability-based multi-path routing strategy, which dynamically adjusts traffic paths based on real-time link loads. This approach enables LGSR to adapt to varying network conditions and distribute traffic more evenly.
\subsection{Experimental Results}
\subsubsection{Load Balancing Performance Analysis}
To comprehensively evaluate the load balancing performance of different algorithms, we conducted three sets of comparative experiments. These experiments simulated various scenarios across six different satellite network scales, six different numbers of flow tunnels, and varying sizes of flow tunnels. Figure \ref{gini} presents the Gini coefficient results for each algorithm under different parameters, providing a measure of load balancing effectiveness.
The results show that our proposed algorithm outperforms both the RAN and Greedy algorithms in terms of load balancing, while only slightly lagging behind the Dijkstra algorithm. Additionally, it can be observed that variations in network scale, flow size, and the number of tunnels have minimal impact on the load balancing capabilities of the different algorithms.
This outcome is primarily due to the lack of a global network perspective in the path selection strategies of RAN and Greedy algorithms, making it challenging for them to effectively balance link loads in large-scale networks. The limitations and randomness of their path selection are further amplified in complex network environments. In contrast, our proposed algorithm integrates global network topology and link load information and adjusts the path selection strategy through probabilistic forwarding, resulting in a more effective load balancing performance across all conditions.
\subsubsection{Latency Analysis and Comparison}
\begin{figure*}[htbp]
    \centering
    \begin{minipage}{0.31\linewidth}
        \centering
        \includegraphics[width=1\linewidth]{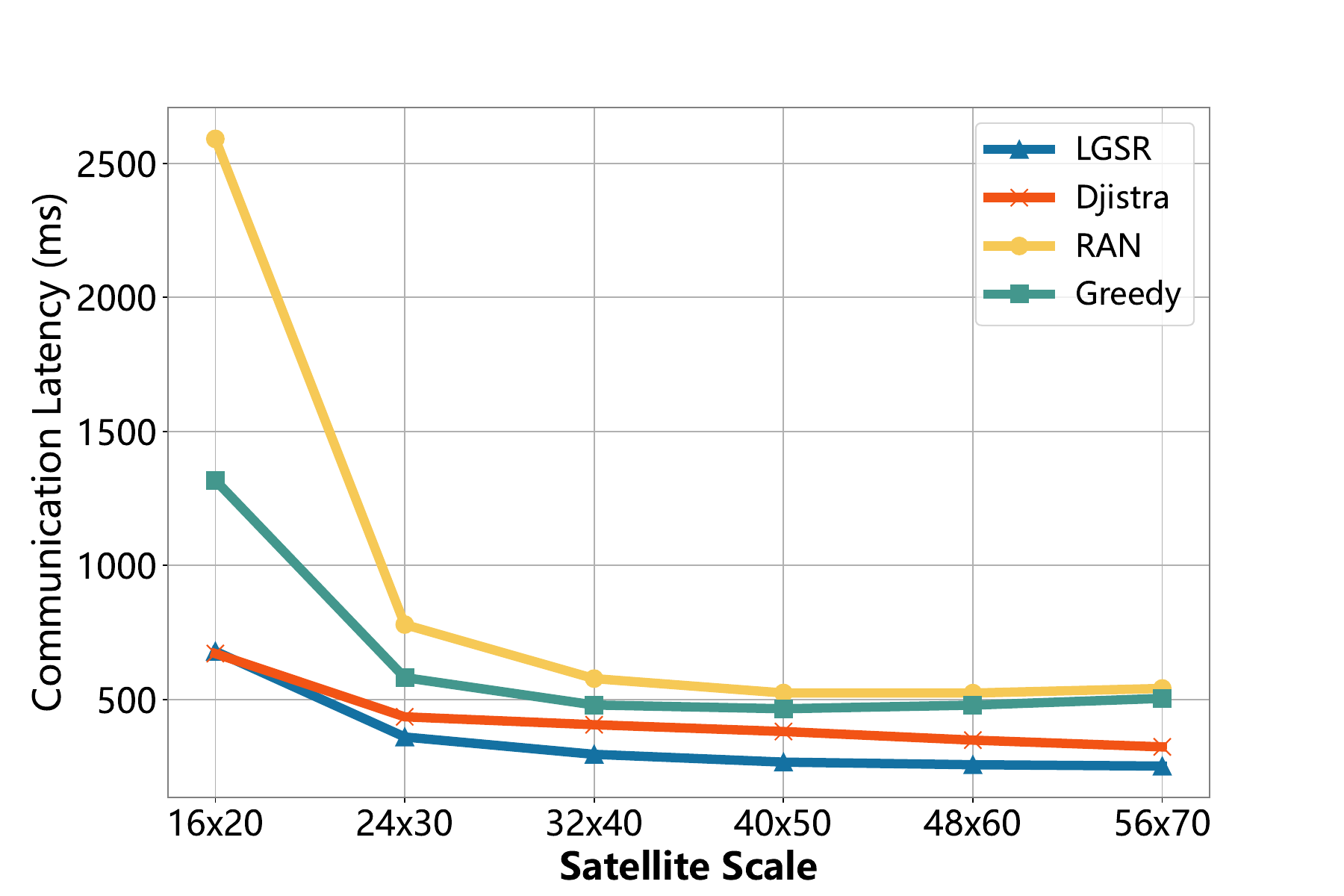}
        \caption*{(\textbf{a}) Communication Delay by Satellite Scale}
        \label{fig:delay_scale}
    \end{minipage}
    \hspace{0.02\linewidth} 
    \begin{minipage}{0.31\linewidth}
        \centering
        \includegraphics[width=1\linewidth]{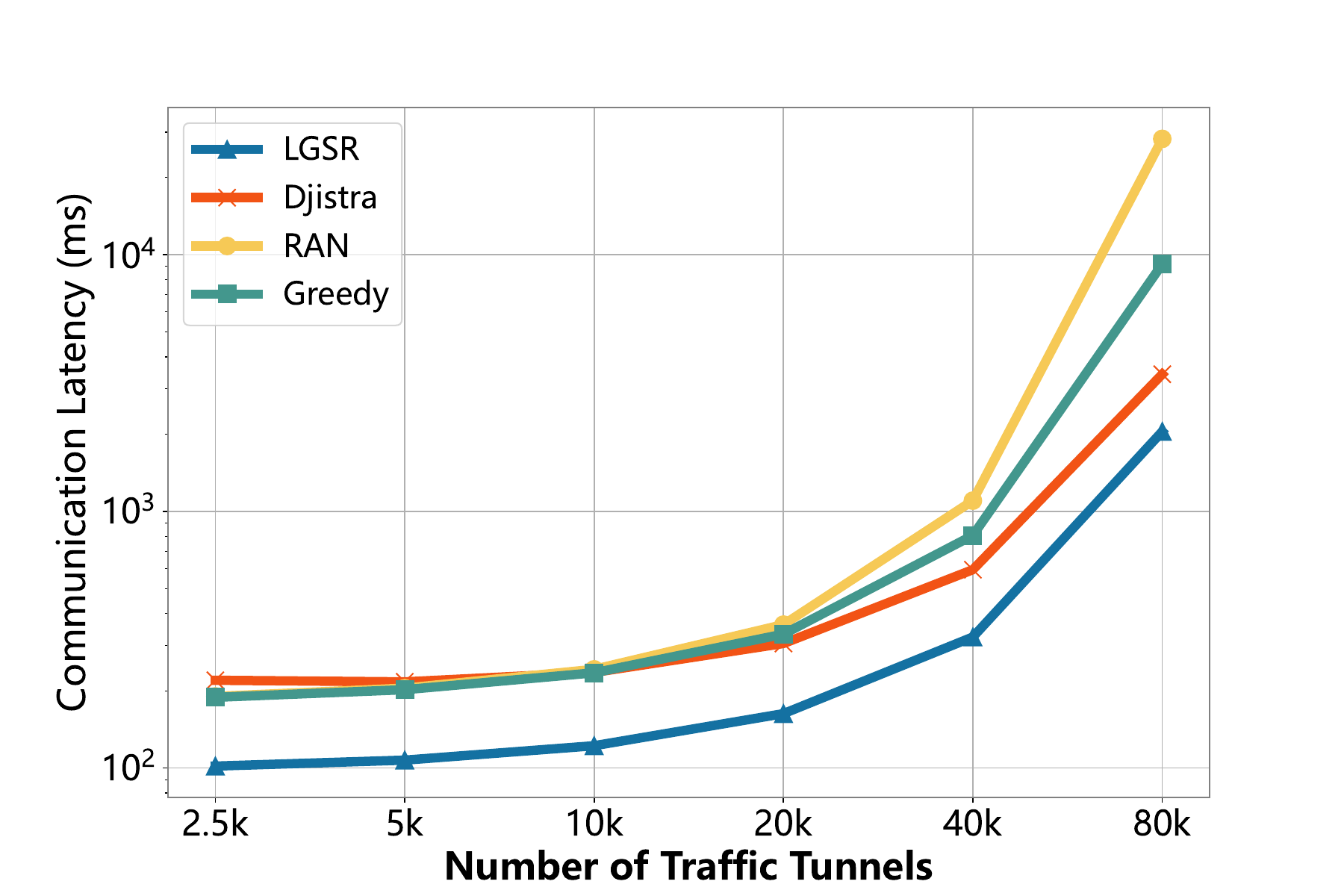}
        \caption*{(\textbf{b}) Communication Delay by Number of Tunnels}
        \label{fig:delay_tunnels}
    \end{minipage}
    \hspace{0.02\linewidth} 
    \begin{minipage}{0.31\linewidth}
        \centering
        \includegraphics[width=1\linewidth]{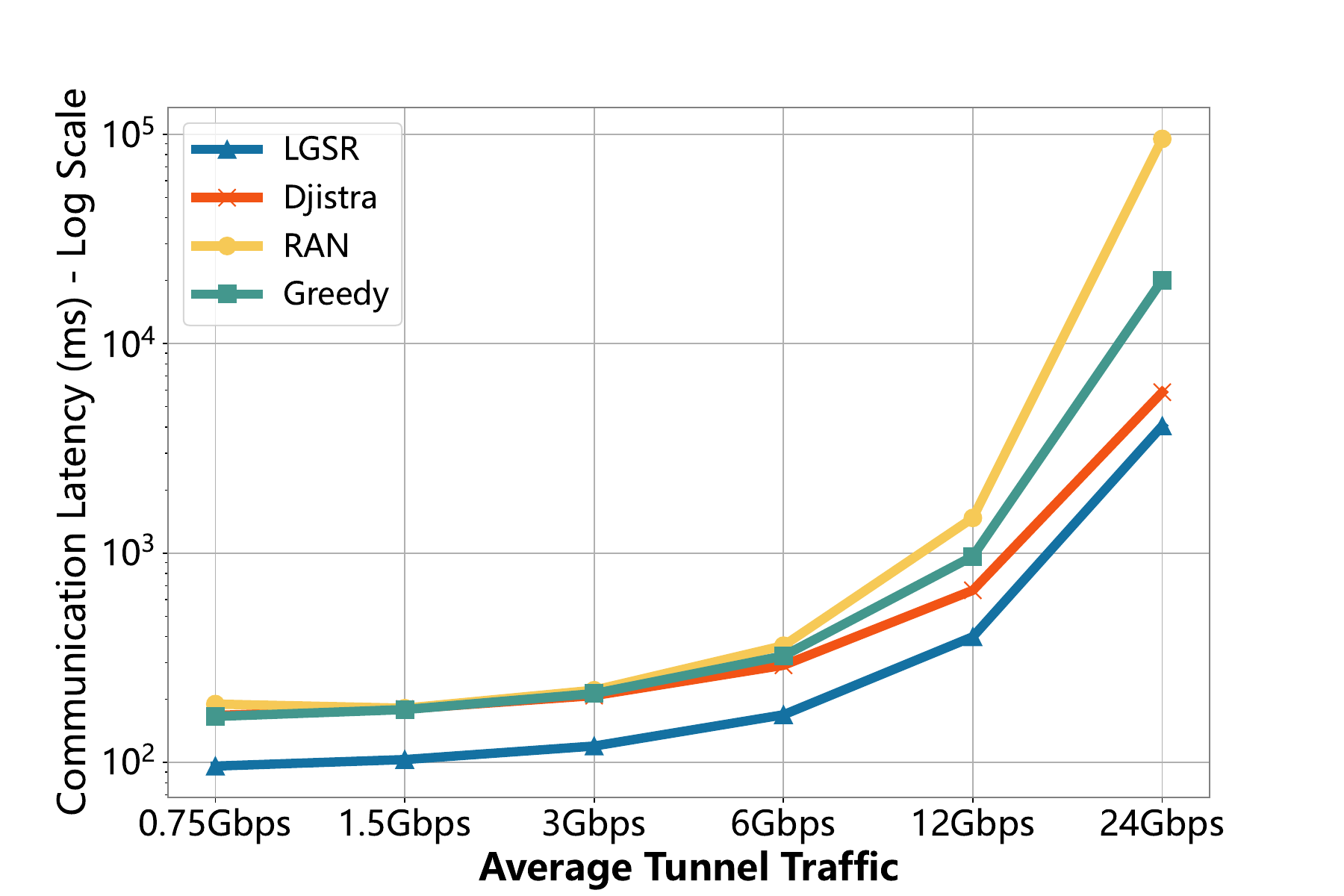}
        \caption*{(\textbf{c}) Communication Delay by Tunnel Traffic Size}
        \label{fig:delay_avg_tunnel}
    \end{minipage}
    \caption{Communication delay under different conditions: (\textbf{a}) by satellite scale, (\textbf{b}) by number of tunnels, and (\textbf{c}) by tunnel traffic size.}
    \label{fig:communication_delay}
\end{figure*}

\begin{figure*}[htbp]
    \centering
    \begin{minipage}{0.31\linewidth}
        \centering
        \includegraphics[width=1\linewidth]{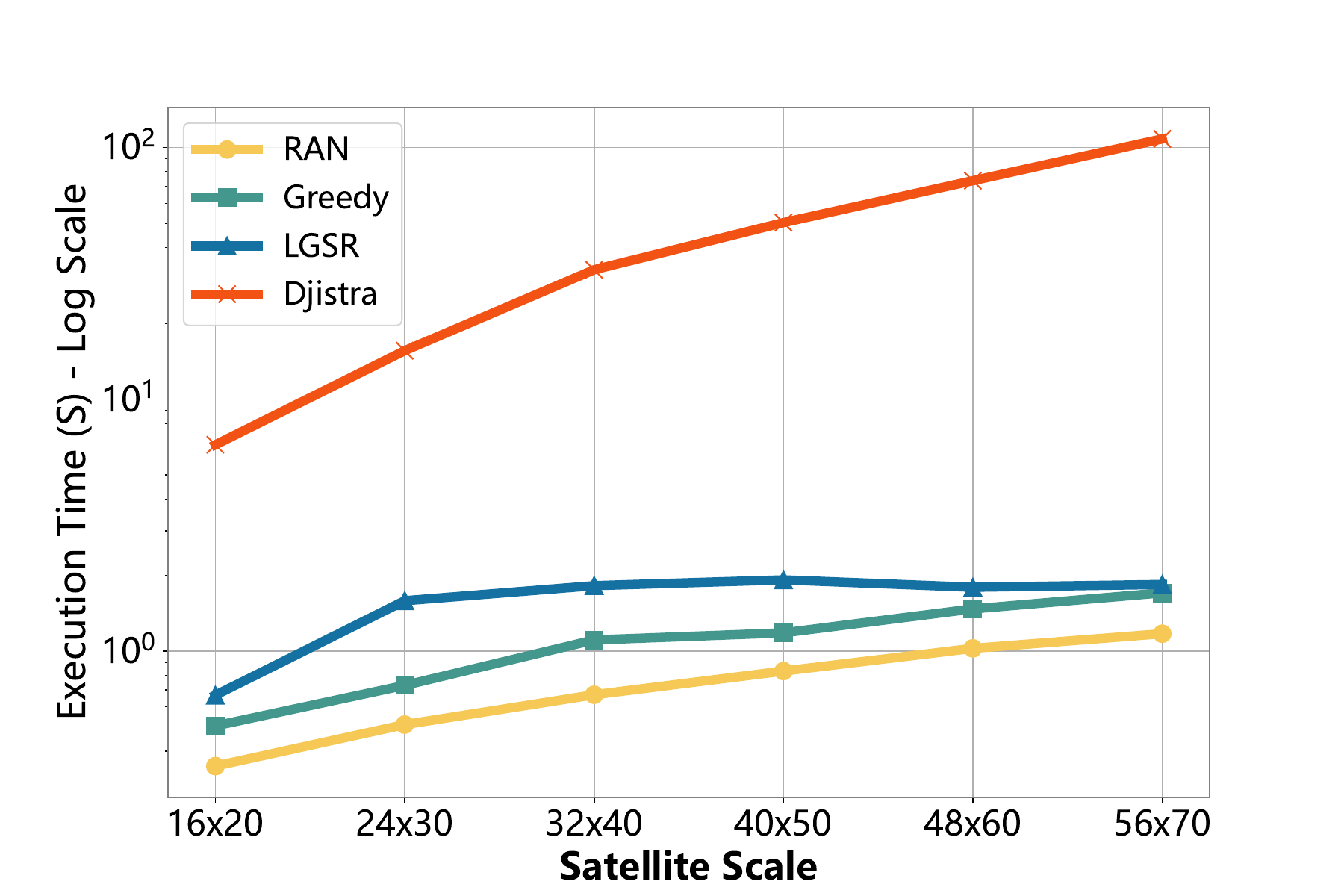}
        \caption*{(\textbf{a}) Execution Time by Satellite Scale}
        \label{fig:exec_time_scale}
    \end{minipage}
    \hspace{0.02\linewidth} 
    \begin{minipage}{0.31\linewidth}
        \centering
        \includegraphics[width=1\linewidth]{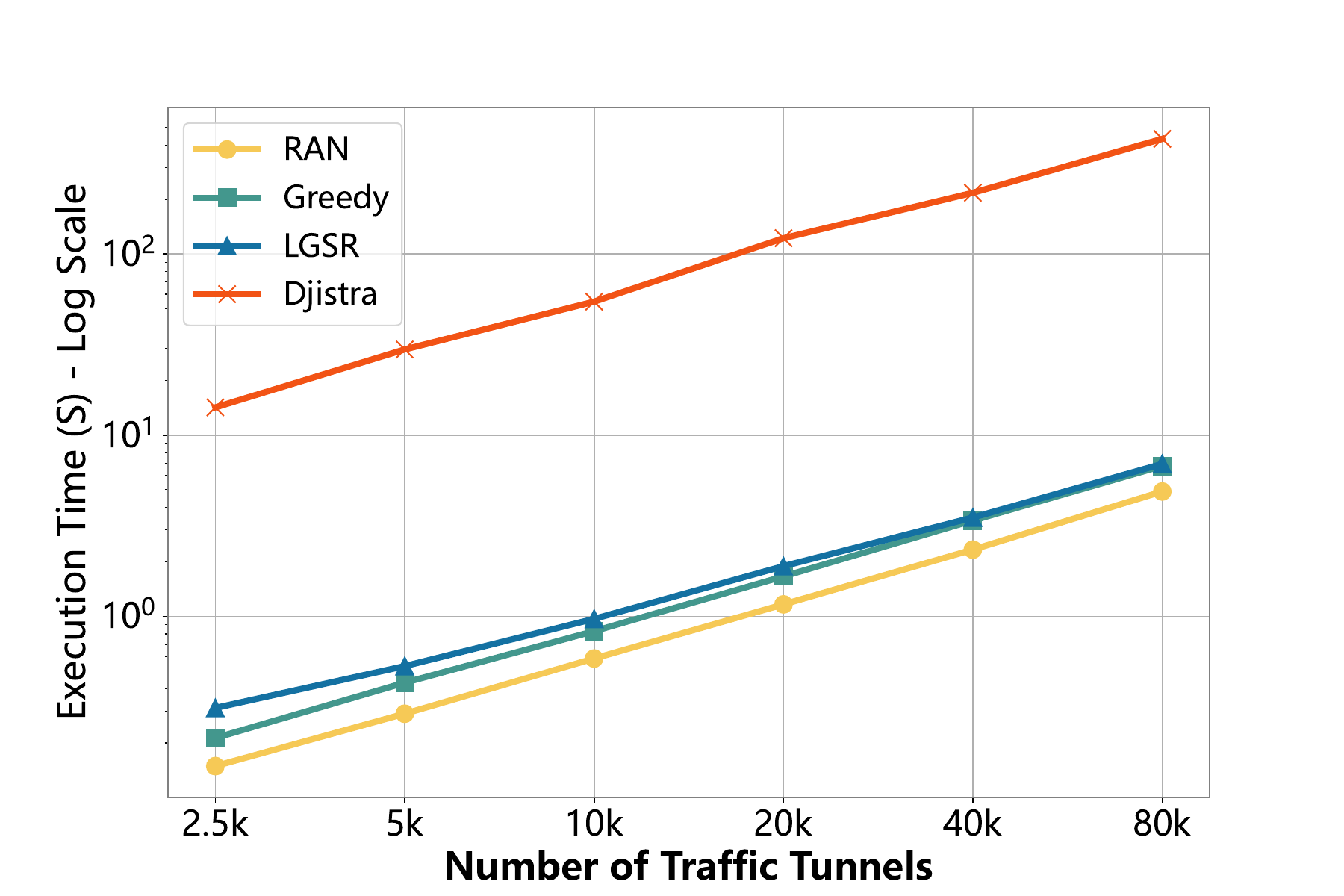}
        \caption*{(\textbf{b}) Execution Time by Number of Tunnels}
        \label{fig:exec_time_tunnels}
    \end{minipage}
    \hspace{0.02\linewidth} 
    \begin{minipage}{0.31\linewidth}
        \centering
        \includegraphics[width=1\linewidth]{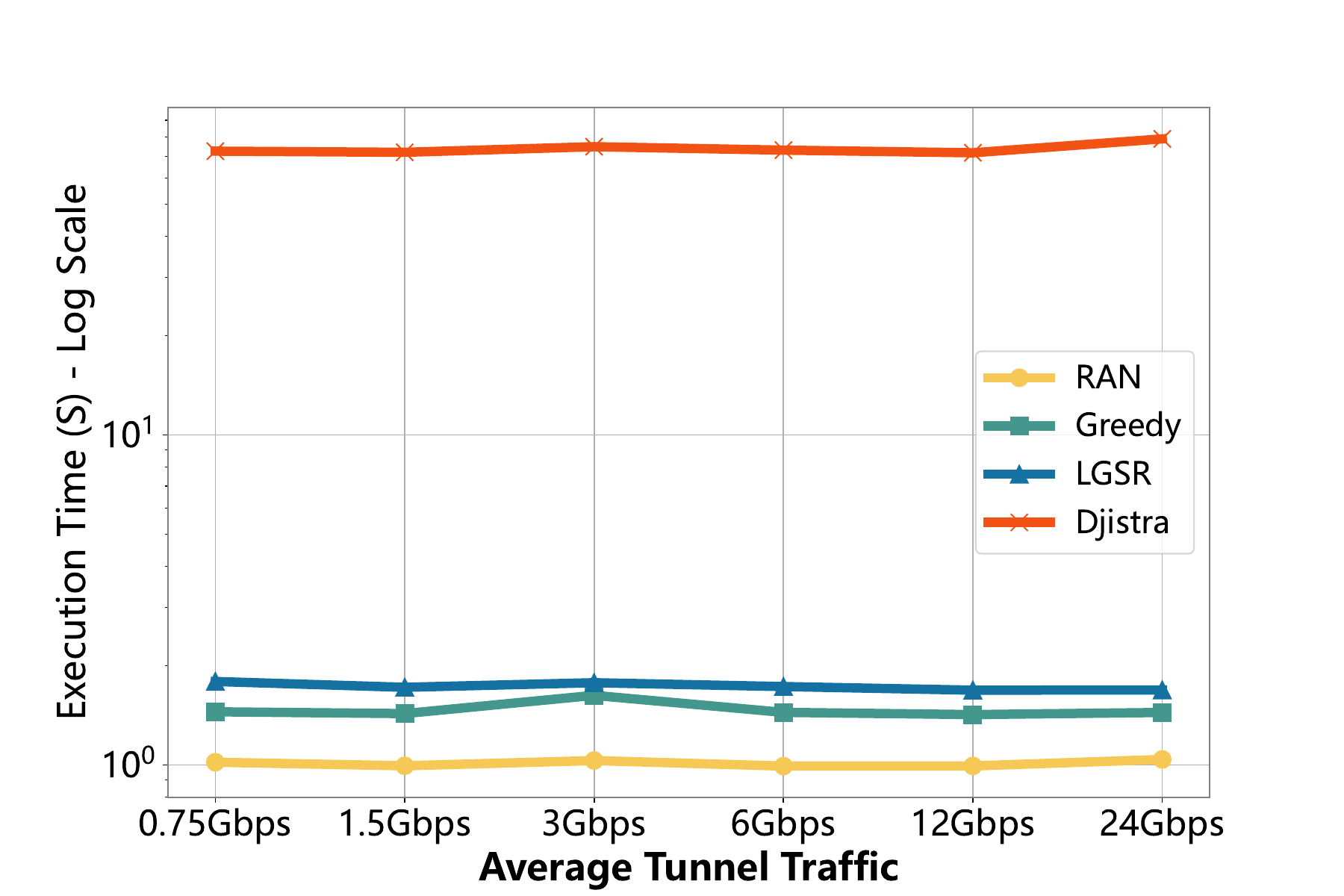}
        \caption*{(\textbf{c}) Execution Time by Tunnel Traffic Size}
        \label{fig:exec_time_avg_tunnel}
    \end{minipage}
    \caption{Algorithm execution time under different conditions: (\textbf{a}) by satellite scale, (\textbf{b}) by number of tunnels, and (\textbf{c}) by tunnel traffic size.}
    \label{fig:algorithm_execution_time}
\end{figure*}

Figure \ref{fig:communication_delay} presents a comparison of the average communication delay for four routing algorithms under varying satellite network scales, different numbers of flow tunnels, and various tunnel sizes. As network scale increases, communication delay decreases across all algorithms, primarily due to reduced link utilization rates in larger networks, which in turn lowers link delay. Across different network scales, our proposed LGSR algorithm consistently maintains low communication delay.
Additionally, as both the number and size of flow tunnels increase, communication delay exhibits a corresponding upward trend. However, LGSR shows a relatively smaller increase in delay compared to other algorithms, demonstrating robust load resistance and the ability to maintain low delay even under higher network load. This effective delay control highlights the advantage of LGSR in managing link loads efficiently.
Although the Dijkstra algorithm, through global shortest path search, theoretically avoids link congestion, it suffers from high computational complexity and less effective delay control in complex network conditions compared to LGSR. By integrating global network topology and link load information, the LGSR algorithm optimizes path selection through probabilistic forwarding, achieving advantages not only in delay control but also in load balancing and scalability.

\subsubsection{Analysis of Algorithm Execution Time}
Figure \ref{fig:algorithm_execution_time} illustrates the execution time (log-scaled \(\log_{10}\)) of four routing algorithms across various satellite network scales, numbers of flow tunnels, and tunnel sizes. The results indicate that the RAN and Greedy algorithms exhibit slightly faster execution times than LGSR, primarily due to their lower computational complexity. Both RAN and Greedy employ local optimization or random strategies for path selection, without requiring global network state or link load information.
In contrast, the Dijkstra algorithm, which relies on a global shortest path search, has a significantly higher computational complexity and demonstrates a clear disadvantage in execution time. As network scale, tunnel count, and tunnel size increase, the execution time of Dijkstra grows exponentially, reaching 108 seconds in a 56×70 network—substantially exceeding the runtime of the other algorithms. LGSR, on the other hand, maintains relatively low execution times even in complex network conditions, requiring only 1.8 seconds to complete route planning.
These findings highlight that LGSR offers substantial advantages in computational efficiency over Dijkstra while maintaining effective load balancing and low communication delay. Although RAN and Greedy algorithms show slight advantages in execution time, their performance in load balancing and communication delay is notably inferior to that of LGSR.
\section{Conclusion}
The paper presents the LGSR algorithm designed to optimize routing in Low Earth Orbit (LEO) satellite networks, effectively addressing critical challenges such as high dynamics, limited resources, and traffic imbalance. LGSR employs a lightweight segment routing approach based on landmark-based skeleton graphs, which significantly enhances the efficiency of path calculation and update processes by reducing computational complexity and transmission costs. 
Moreover, the algorithm demonstrates effective management of communication delay, maintaining low latency even under increased network loads, while ensuring robust load balancing across the network.
Experimental results validate that LGSR outperforms traditional routing algorithms, including Dijkstra, in terms of both communication delay and execution time. Specifically, LGSR achieves route planning in merely 1.8 seconds in larger networks, in stark contrast to Dijkstra's 108 seconds. Although the RAN and Greedy algorithms exhibit faster execution times, they do not match LGSR's efficiency in load balancing and delay control.

\bibliographystyle{IEEEtran}
\bibliography{LGSR_ref}

\end{document}